\documentclass[12pt]{article}
\usepackage{url}

\usepackage{multirow}
\usepackage{diagbox}
\usepackage{bm}
\usepackage{lineno} 
\usepackage{amsfonts}
\usepackage{amssymb}
\usepackage{amsmath}
\usepackage{amsthm}
\usepackage{amssymb}
\usepackage{algorithm}
\usepackage{algpseudocode}
\usepackage{latexsym}
\usepackage{graphicx,color}
\usepackage{graphics}
\usepackage{epsfig}
\usepackage{psfrag}
\usepackage{threeparttable}
\usepackage{booktabs}
\usepackage{lineno}
\theoremstyle{plain}

\usepackage{color}
\definecolor{darkred}{rgb}{0.62,0.18,0.18}

\usepackage{longtable,lscape}
\numberwithin{equation}{section}
\begin{document}
\fontsize{12}{25pt}\selectfont
\title{\textbf{Efficient estimation for flexible spatial zero-inflated models with environmental applications}}
\author{
    {{Chung-Wei Shen$^{1}$, Bu-Ren Hsu$^{2}$, Chia-Ming Hsu$^{2}$, and Chun-Shu Chen}$^{2}$\thanks{Corresponding author. Tel.: +886-3-4227151 ext. 65456; Email: cschen1207$@$ncu.edu.tw}}\\ \\
    \small $^{1}$Department of Mathematics, National Chung Cheng University, Chia-Yi, Taiwan, R.O.C.
    \and
    \small $^{2}$Graduate Institute of Statistics, National Central University, Taoyuan, Taiwan, R.O.C.
}
\date{}
\maketitle
\begin{abstract}
\fontsize{12}{22pt}\selectfont
   Spatial two-component mixture models offer a robust framework for analyzing spatially correlated data with zero inflation. To circumvent potential biases introduced by assuming a specific distribution for the response variables, we employ a flexible spatial zero-inflated model. Despite its flexibility, this model poses significant computational challenges, particularly with large datasets, due to the high dimensionality of spatially dependent latent variables, the complexity of matrix operations, and the slow convergence of estimation procedures. To overcome these challenges, we propose a projection-based approach that reduces the dimensionality of the problem by projecting spatially dependent latent variables onto a lower-dimensional space defined by a selected set of basis functions. We further develop an efficient iterative algorithm for parameter estimation, incorporating a generalized estimating equation (GEE) framework. The optimal number of basis functions is determined using Akaike's information criterion (AIC), and the stability of the parameter estimates is assessed using the block jackknife method. The proposed method is validated through a comprehensive simulation study and applied to the analysis of Taiwan's daily rainfall data for 2016, demonstrating its practical utility and effectiveness.
\end{abstract}


\section{Introduction}\label{sec1:introduction}
 Spatial data with a significant number of zeros and spatial dependence is referred to as zero-inflated spatially correlated data. These datasets are prevalent in various fields, such as ecology and atmospheric science. For example, count data for isopod nest burrows in the Negev desert (Agarwal et al. 2002) and semi-continuous data for torrential rainfall events in specific regions (Lee and Kim 2017) often exhibit excess zeros. These data typically consist of a mixture of zeros and either discrete counts or positive real numbers, which standard probability models fail to adequately address due to the overdispersion caused by the excess zeros (Lambert 1992; Rathbun and Fei 2006). To address these challenges, various spatial zero-inflated models have been developed. Early models, such as the zero-inflated Poisson (ZIP) model (Lambert 1992), incorporated a two-component mixture structure to handle excess zeros effectively. Extensions to spatial contexts were introduced by Agarwal et al. (2002) and Rathbun and Fei (2006), who modeled spatial dependence through covariates or latent spatial processes. These foundational models provide flexibility and robustness but often rely on strong parametric assumptions. Moverover, Bayesian approaches further extended spatial zero-inflated models by integrating spatial random effects. Neelon et al. (2015) proposed a Bayesian spatial zero-inflated negative binomial model that utilized Gaussian process priors to capture spatial correlation. This method enhances estimation robustness, particularly in cases with complex spatial structures or limited sample sizes.

 In practical applications, standard probability distributions such as Binomial, Negative Binomial, or Poisson distributions are often employed to construct spatial zero-inflated models (or spatial two-component mixture models). However, the risk of distributional misspecification can lead to erroneous statistical inference, particularly when the underlying data structure deviates from assumed parametric forms. To address this, semiparametric and nonparametric models have gained traction as flexible alternatives. For example, He et al. (2015) combined a distribution-free technique with zero-inflated structures to flexibly model spatial dependencies, mitigating biases associated with parametric assumptions. Building on this foundation, Shen and Chen (2024) proposed a semiparametric spatial zero-inflated count model that introduces two spatially dependent latent variables to separately characterize the occurrence of zeros and non-zeros in the responses. This approach, which relaxes strict parametric assumptions to enhance the model’s adaptability across diverse data distributions and improve its robustness for analyzing spatially correlated data with an excess of zeros, will also be adopted in this study.

 In the era of big data, analyzing large zero-inflated spatial datasets poses significant challenges from both computational and modeling perspectives. First, fitting spatial two-component models can be computationally burdensome for large datasets due to intensive matrix operations and slow convergence of estimation procedures. Second, the underlying spatially dependent latent variables may exhibit complex spatial dependence characteristics, such as non-stationarity and anisotropy, which are difficult to anticipate and incorporate into the model beforehand.

 To address these issues, we designed a flexible and computationally efficient approach for handling high-dimensional matrix operations. This approach uses a projection-based technique to project the original spatially dependent latent variables onto a low-dimensional space spanned by a set of basis functions, while still preserving the inherent structure of the original data (e.g., Cressie and Johannesson 2008; Nychka et al. 2015; Katzfuss 2017). To our knowledge, this is the first time basis functions have been integrated with a spatial zero-inflated model to address computational issues in large datasets. Importantly, this method does not require specifying a parametric covariance structure for spatially dependent latent variables, offering more flexibility in spatial modeling.

 Moreover, there are two important quantities in the flexible zero-inflated spatially correlated model: occurrence probabilities and prevalence intensities. These quantities vary with respect to space and are modeled by generalized linear models with possible spatially varying covariates. Since there are no specific distribution assumptions on the spatial zero-inflated model, we plan to estimate the regression parameters of the generalized linear models using the generalized estimating equation (GEE) approach of Liang and Zeger (1986).

 In this approach, our proposed projection-based technique will be used to construct the covariance matrix of spatially dependent latent variables as the working covariance matrix in GEE. An efficient iterative algorithm will then be proposed to estimate all the parameters in the flexible zero-inflated spatially correlated model.

 In our approach, the number of basis functions for the projection-based technique is determined using the extended Akaike's information criterion (AIC) proposed by Tzeng and Huang (2018). Additionally, a block jackknife (BJ) method, similar to that employed by Adegboye et al. (2018), is used to assess the stability of our proposed estimators. Numerical studies indicate that the performance of our estimation method is satisfactory across various simulation scenarios. For practical applications, we analyze a real data example concerning Taiwan's daily rainfall data for 2016 to demonstrate the feasibility of our proposed methodology.

 The remainder of this approach is structured as follows. In Section \ref{sec2:spatial zero-inflated models}, we introduce the flexible zero-inflated spatially correlated model for discrete data. Section \ref{sec3:Dimension reduction for spatial covariance structures} provides an overview of thin-plate spline functions, detailing their use in constructing the covariance matrix of spatially dependent latent variables through extracted basis functions to facilitate dimensionality reduction. In Section \ref{sec4:estimation of model parameters}, we outline an iterative algorithm for estimating model parameters and introduce the block jackknife method for assessing the variances of the proposed estimators. Sections \ref{sec5:simulation study} and \ref{sec6:application} discuss the simulation results under various scenarios and analyze a real dataset concerning Taiwan's daily rainfall in 2016, respectively. Finally, the paper concludes with Section \ref{sec7:conclusion and discussion}, offering a discussion of the findings and implications of the research. Additional technical details are provided in the online Supplementary Materials.

\section{Flexible spatial zero-inflated models}\label{sec2:spatial zero-inflated models}
 In this section, we introduce how spatial zero-inflated models effectively characterize datasets with an excessive number of zeros. Let the study region $\mathcal{D}\subset\mathbb{R}^2$ be continuous and bounded. We denote the vector $\bm{Y} = (Y(\bm{s}_1), \dots, Y(\bm{s}_n))^t$ as the count response variables observed at $n$ distinct locations $\bm{s}_1, \dots, \bm{s}_n \in \mathcal{D}$ from a spatial random field of interest, with $\bm{s}_i = (s_{i,1}, s_{i,2})^t$ for $i = 1, \dots, n$. Associated with the response variable $Y(\bm{s}_i)$ for $i = 1, \dots, n$, there are $p$ covariates collected at location $\bm{s}_i \in \mathcal{D}$, denoted along with a 1 for the intercept, as $\bm{x}(\bm{s}_i) = (1, x_1(\bm{s}_i), \dots, x_p(\bm{s}_i))^t$. Furthermore, we denote $\bm{X} = (\bm{x}(\bm{s}_1), \dots, \bm{x}(\bm{s}_n))^t$ as a non-random $n \times (p+1)$ design matrix of full column rank.

 For a location $\bm{s}_{i} \in \mathcal{D}$, where $i = 1, \ldots, n$, consider a random variable $Y^*(\bm{s}_{i})$ defined on the set of non-negative integers, $\mathcal{A}=\{0, 1, 2, \ldots\}$, which represents the count of occurrences of events. The mean rate of these events, $\lambda(\bm{s}_{i})$, is positive and varies with $\bm{s}_{i}$. This variable can be appropriately modeled using a Poisson distribution as described below:
 \begin{equation}\label{poisson pdf}
  P\big(Y^*(\bm{s}_{i}) = y(\bm{s}_{i}) \mid \lambda(\bm{s}_{i})\big) =
  \frac{e^{-\lambda(\bm{s}_{i})} \lambda(\bm{s}_{i})^{y(\bm{s}_{i})}}{y(\bm{s}_{i})!}, \quad y(\bm{s}_{i})\in\mathcal{A}.
 \end{equation}

 To model the count response $Y(\bm{s}_i)$ at locations $\bm{s}_i \in \mathcal{D}$, which exhibits an excessive number of zeros, we utilize the formulation presented in \eqref{poisson pdf}, referred to as the zero-inflated Poisson (ZIP) distribution. This distribution has been proposed and extensively employed across various scientific domains (e.g., Farewell and Sprott 1988; B$\ddot{\rm{o}}$hning et al. 1999; Silva et al. 2014). Here, the ZIP distribution is defined as follows:
 \begin{align}\label{zip pdf}
  &\widetilde{P}\big(Y(\bm{s}_{i})=y(\bm{s}_{i}) \mid \lambda(\bm{s}_{i}), \phi(\bm{s}_{i})\big)\nonumber\\
  &=
  \begin{cases}
  \phi(\bm{s}_{i}) + (1 - \phi(\bm{s}_{i}))P\big(Y^*(\bm{s}_{i}) = y(\bm{s}_{i}) \mid \lambda(\bm{s}_{i})\big), & \text{if } y(\bm{s}_{i}) = 0, \\
  (1 - \phi(\bm{s}_{i}))P\big(Y^*(\bm{s}_{i}) = y(\bm{s}_{i}) \mid \lambda(\bm{s}_{i})\big), & \text{if } y(\bm{s}_{i})\in\mathcal{A}\backslash\{0\}.
  \end{cases}
 \end{align}

 The ZIP model, as defined in \eqref{zip pdf}, can be conceptualized as a two-component mixture model. It incorporates a location-dependent mixture probability $\phi(\bm{s}_{i})$ to account for the excess zeros observed in the data. In this paper, both $\phi(\bm{s}_{i})$ and $\lambda(\bm{s}_{i})$ are modeled using generalized linear models. The respective forms of these models are defined as follows:
 \begin{equation}\label{phi}
  \log\Bigg(\frac{\phi(\bm{s}_{i})}{1 - \phi(\bm{s}_{i})}\Bigg) = \bm{v}(\bm{s}_{i})^{t} \bm{\beta}
 \end{equation}
 and
 \begin{equation}\label{lambda}
  \log(\lambda(\bm{s}_{i})) = \bm{u}(\bm{s}_{i})^{t} \bm{\gamma},
 \end{equation}
 where $\bm{v}(\bm{s}_{i}) \subseteq \bm{x}(\bm{s}_{i})$ and $\bm{u}(\bm{s}_{i}) \subseteq \bm{x}(\bm{s}_{i})$ are covariates for $i = 1, \ldots, n$, and $\bm{\beta}$ and $\bm{\gamma}$ are the corresponding unknown vectors of regression coefficients associated with these covariates.

 In practice, the underlying distribution of data $\bm{Y}$ is unknown. To avoid making incorrect inferences by imposing a specific distribution on the observed count variable $\bm{Y}$, we adopt the approach of Shen and Chen (2024). This approach separately models the probability of structural zeros (i.e., $Y(\bm{s}_i)=0$) and the mean of a zero-truncated Poisson random variable (i.e., $Y(\bm{s}_i)>0$) as given in \eqref{zip pdf}. Consequently, we have:

 \begin{equation}\label{zero mean}
  E[I(Y(\bm{s}_{i}) = 0) \mid \bm{x}(\bm{s}_{i})] = \phi(\bm{s}_{i}) + (1 - \phi(\bm{s}_{i}))e^{-\lambda(\bm{s}_{i})}
 \end{equation}
 and
 \begin{equation}\label{nonzero mean}
  E[Y(\bm{s}_{i}) \mid Y(\bm{s}_{i}) > 0, \bm{x}(\bm{s}_{i})] = \frac{\lambda(\bm{s}_{i})}{1 - e^{-\lambda(\bm{s}_{i})}},
 \end{equation}
 where $I(\cdot)$ is an indicator function. It is evident that models \eqref{zero mean} and \eqref{nonzero mean} can be derived from model \eqref{zip pdf}, but the converse is not necessarily true. By using models \eqref{zero mean} and \eqref{nonzero mean}, we can model zero-inflated count data $\bm{Y}$ without relying on the exact distributional assumptions specified in model \eqref{zip pdf}, offering a flexible approach that provides more robust inferences in practical applications. In this paper, we refer to models \eqref{zero mean} and \eqref{nonzero mean} as a flexible spatial zero-inflated count model that does not rely on full distributional assumptions. While the spatial hurdle model, widely applied to zero-inflated data, handles a mixture of zeros and truncated (positive) counts (e.g., Neelon et al. 2012; Chen and Chen 2024), it does not differentiate between structural zeros and random zeros. This limitation restricts its ability to provide inference on the possibility of structural zeros, an important aspect in many practical contexts. In contrast, our proposed model explicitly incorporates structural zeros (e.g., via Equation \eqref{zero mean}) and allows for the differentiation between structural and random zeros, providing a significant advantage in settings where understanding the nature of zero counts is critical. By explicitly modeling structural zeros and avoiding strong distributional assumptions, our approach enhances flexibility and robustness, particularly in applications where the true data-generating process may deviate from parametric models. This flexibility in modeling data-generating mechanisms reflects a hybrid modeling strategy that combines parametric structure (e.g., moment specification) with nonparametric adaptability.

 To account for the spatial dependence of data $\bm{Y}$, we introduce two latent random variables, $Z_{1}(\bm{s}_{i})$ and $Z_{2}(\bm{s}_{i})$, for $i = 1, \ldots, n$. These variables characterize the occurrence of zeros and non-zeros in $\bm{Y}$, respectively, based on models \eqref{zero mean} and \eqref{nonzero mean}. Specifically, we define $Z_{1}(\bm{s}_{i})$ and $Z_{2}(\bm{s}_{i})$ as follows:
 \begin{align}\label{rv for zero(zip-like)}
  Z_{1}(\bm{s}_{i})
  &=
  I(Y(\bm{s}_{i}) = 0) - E[I(Y(\bm{s}_{i}) = 0)| \bm{x}(\bm{s}_{i})]
  \notag \\
  &=
  I(Y(\bm{s}_{i}) = 0) - \phi(\bm{s}_{i}) - (1 - \phi(\bm{s}_{i}))e^{-\lambda(\bm{s}_{i})}
 \end{align}
 and
 \begin{align}\label{rv for nonzero(zip-like)}
  Z_{2}(\bm{s}_{i})
  &=
  I(Y(\bm{s}_{i}) > 0)\Big(Y(\bm{s}_{i}) - E[Y(\bm{s}_{i})|Y(\bm{s}_{i}) > 0, \bm{x}(\bm{s}_{i})]\Big)
  \notag \\
  &=
  I(Y(\bm{s}_{i}) > 0)\left( Y(\bm{s}_{i}) - \frac{\lambda(\bm{s}_{i})}{1 - e^{-\lambda(\bm{s}_{i})}} \right)
 \end{align}
 for $i = 1, \ldots, n$. We then obtain the variances of $Z_{1}(\bm{s}_{i})$ and $Z_{2}(\bm{s}_{i})$ for $i = 1, \ldots, n$, as follows:
 \begin{align}
  \text{Var}\big(Z_{1}(\bm{s}_{i})\big)
  &=
  \Big[\phi(\bm{s}_{i}) + (1 - \phi(\bm{s}_{i}))e^{-\lambda(\bm{s}_{i})}\Big]
  \Big[\big( 1 - \phi(\bm{s}_{i}) \big) \Big(1 - e^{-\lambda(\bm{s}_{i})}\Big)\Big]
  \label{var rv for zero(zip-like)}
 \end{align}
 and
 \begin{align}
  \text{Var}\big(Z_{2}(\bm{s}_{i})\big)
  &=
  \Big[\big(1 - \phi(\bm{s}_{i}) \big)\Big(1 - e^{-\lambda(\bm{s}_{i})} \Big)\Big]
  \left[ \frac{\lambda(\bm{s}_{i}) + \lambda^2(\bm{s}_{i})}{1 - e^{-\lambda(\bm{s}_{i})}} - \left( \frac{\lambda(\bm{s}_{i})}{1 - e^{-\lambda(\bm{s}_{i})}}\right)^2 \right].
  \label{var rv for nonzero(zip-like)}
 \end{align}
 In addition, we can show that $\text{Cov}\big(Z_{1}(\bm{s}_{i}), Z_{2}(\bm{s}_{j}) \big) = 0$ for $i, j = 1, \ldots, n$. The derivation of the variances and covariance for $Z_{1}(\bm{s}_{i})$ and $Z_{2}(\bm{s}_{i})$ can be found in the online Supplementary Materials. However, verifying the magnitude of $\text{Cov}\big( Z_{k}(\bm{s}_{i}), Z_{k}(\bm{s}_{j}) \big)$ for $k = 1, 2$ and $i, j = 1, \ldots, n$, with $i \neq j$, is highly challenging. Therefore, we address this issue using the technique of thin-plate splines (TPS) (e.g., Wahba 1985; 1990) as introduced in Section \ref{sec3.1:ApplyTPS}.

\section{Dimension reduction for covariance structures}\label{sec3:Dimension reduction for spatial covariance structures}
 In this section, we discuss the use of basis functions derived from thin-plate splines (TPS) (e.g., Wahba 1985; 1990) to construct the spatial covariance structure of the data $\bm{Y}$. First, we define $\bm{Z}_{1} = (Z_{1}(\bm{s}_{1}), \ldots, Z_{1}(\bm{s}_{n}))^t$, $\bm{Z}_{2} = (Z_{2}(\bm{s}_{1}), \ldots, Z_{2}(\bm{s}_{n}))^t$, and $\bm{Z} = (\bm{Z}_{1}^t, \bm{Z}_{2}^t)^t$. We then denote the variance of $\bm{Z}$, given the covariates $\bm{X}$, as $\bm{\Sigma}_{K_{1}, K_{2}} \equiv \text{Var}(\bm{Z}|\bm{X})$, where $K_1$ and $K_2$ are two unknown parameters. Next, we provide details on constructing $\bm{\Sigma}_{K_{1}, K_{2}}$ using basis functions and then introduce the basis functions derived from TPS.

\subsection{Constructing covariance matrices}\label{sec3.1:ApplyTPS}
 In the era of big data, analyzing large zero-inflated spatial datasets presents significant challenges from both computational and modeling perspectives. First, fitting spatial two-component models to large datasets can be computationally intensive due to operations involving the large matrices. Second, the underlying spatially dependent latent variables might exhibit complex spatial dependence patterns, such as non-stationarity and anisotropy, which are difficult to anticipate and incorporate into the model beforehand. To address these challenges, we employ a flexible and computationally efficient method to handle high-dimensional covariance matrices (i.e., $\bm{\Sigma}_{K_{1}, K_{2}} \equiv \text{Var}(\bm{Z}|\bm{X})$) using the basis functions.

 Consider the process $\{Z_j(\bm{s}): \bm{s} \in \mathcal{D}\}$ defined on a spatial domain $\mathcal{D}$, with zero mean and a covariance function $\gamma_j(\bm{s}, \bm{s}^*) = \text{Cov}(Z_j(\bm{s}), Z_j(\bm{s}^*))$, for $j = 1, 2$ and $\bm{s}, \bm{s}^* \in \mathcal{D}$. Let $\bm{Z}_j = (Z_j(\bm{s}_1), \dots, Z_j(\bm{s}_n))^t$ denote the vector of spatial random variables, where each element has zero mean and the covariance matrix $\bm{\Sigma}_j$ of $\bm{Z}_j$ is defined through the covariance function $\gamma_j$. To model the complex spatial covariance function $\gamma_j$ of $\bm{Z}_j$ (e.g., non-stationarity or anisotropy) for $j = 1, 2$, we adopt a spatial random effects model (e.g., Wikle 2010) for $\{Z_j(\bm{s}): \bm{s} \in \mathcal{D}\}$ as follows:
 \begin{eqnarray}
  Z_j(\bm{s})&=&\bm{\omega}^t_j\bm{\psi}_j(\bm{s})+\xi_j(\bm{s})\nonumber\\
   &=&\displaystyle\sum_{k=1}^{K_j}\omega_{j,k}\psi_{j,k}(\bm{s})+\xi_j(\bm{s}),
  \label{eq:SRE.model}
 \end{eqnarray}
 where $\bm{\omega}_j = (\omega_{j,1}, \dots, \omega_{j,K_j})^t \sim N(\bm{0}, \bm{\Omega}_j)$ represents a vector of random effects with $\bm{\Omega}_j$ as an unknown nonnegative-definite matrix. The functions $\{\psi_{j,k}(\cdot): k = 1, \dots, K_j\}$ are prespecified basis functions with $K_j \leq n$, as given in \eqref{tps basis function}, where $K_j$ denotes the number of basis functions for $j = 1, 2$.
 Here, $\bm{\psi}_j(\bm{s}) = (\psi_{j,1}(\bm{s}), \dots, \psi_{j,K_j}(\bm{s}))^t$, and $\xi_j(\bm{s}) \sim N(0, \sigma^2_{\xi,j})$ is a white-noise process which is independent of $\bm{\omega}_j$. Then, the covariance function $\gamma_j(\bm{s}, \bm{s}^*)$ can be represented as follows:
 \begin{eqnarray}
  \gamma_j(\bm{s},\bm{s}^*)&=&Cov(Z_j(\bm{s}),Z_j(\bm{s}^*))\nonumber\\
  &=&\bm{\psi}^t_j(\bm{s})\bm{\Omega}_j\bm{\psi}_j(\bm{s}^*)+\sigma^2_{\xi,j}I(\bm{s},\bm{s^*}),
  \label{eq:cov.function}
 \end{eqnarray}
 where $I(\bm{s}, \bm{s^*})$ is defined as $1$ when $\bm{s} = \bm{s}^*$ and $0$ when $\bm{s} \neq \bm{s}^*$. Let $\bm{\Psi}_j$ be an $n \times K_j$ matrix with the $(i, k)$-th element being $\psi_{j,k}(\bm{s}_i)$. Then, the covariance matrix of $\bm{Z}_j$ is given by:
 \begin{eqnarray}
  \bm{\Sigma}_{K_j}=\bm{\Psi}_j\bm{\Omega}_j\bm{\Psi}^t_j+\sigma^2_{\xi,j}\bm{I}.
  \label{eq:Z_j.cov.function}
 \end{eqnarray}
 In \eqref{eq:Z_j.cov.function}, the rank of $\bm{\Psi}_j\bm{\Omega}_j\bm{\Psi}^t_j$ does not exceed $K_j$ (i.e., ${\rm{rank}}(\bm{\Psi}_j\bm{\Omega}_j\bm{\Psi}^t_j) \leq K_j$). As a result, this procedure is termed fixed rank kriging (see Cressie and Johannesson 2008). This method notably avoids the need to specify a parametric covariance structure for the spatial random effects, offering greater flexibility in modeling spatial covariance functions. Furthermore, it can be shown that $\text{Cov}(Z_1(\bm{s}), Z_2(\bm{s}^*)) = 0$ for all $\bm{s}, \bm{s}^* \in \mathcal{D}$. Consequently, the covariance matrix of $\bm{Z}$ can be expressed as follows:
 \begin{eqnarray}
  \bm{\Sigma}_{K_1,K_2} =
  \left(
  \begin{array}{cc}
   \bm{\Sigma}_{K_1} & \bm{O} \\
   \bm{O} & \bm{\Sigma}_{K_2} \\
  \end{array}
  \right)
  \label{eq:Z.covariance.matrix}
 \end{eqnarray}
 and the inverse matrix of $\bm{\Sigma}_{K_1,K_2}$ is
 \begin{eqnarray}
  \bm{\Sigma}^{-1}_{K_1,K_2} =
  \left(
  \begin{array}{cc}
   \bm{\Sigma}^{-1}_{K_1} & \bm{O} \\
   \bm{O} & \bm{\Sigma}^{-1}_{K_2} \\
  \end{array}
  \right).
  \label{eq:Inverse.Z.covariance.matrix}
 \end{eqnarray}
 Utilizing the Sherman-Morrison-Woodbury formula (see Harville 1997 for reference), we can provide an alternative expression for $\bm{\Sigma}^{-1}_{K_j}$ ($j = 1, 2$), as presented in equation (\ref{eq:Sherman-Morrison-Woodbury}), to overcome the challenge of high-dimensional matrix operations.
 \begin{eqnarray}
  \bm{\Sigma}^{-1}_{K_j}=\displaystyle\frac{1}{\sigma^2_{\xi,j}}\bm{I}
  -\displaystyle\frac{1}{\sigma^4_{\xi,j}}\bm{\Psi}_j
  \left(\bm{\Omega}^{-1}_j+\frac{1}{\sigma^2_{\xi,j}}\bm{\Psi}^t_j\bm{\Psi}_j\right)^{-1}\bm{\Psi}^t_j.
  \label{eq:Sherman-Morrison-Woodbury}
 \end{eqnarray}
 It is worth noting that equation (\ref{eq:Sherman-Morrison-Woodbury}) demonstrates that the computational dimension of $\bm{\Sigma}^{-1}_{K_j}$ can be effectively reduced from $n \times n$ to $K_j \times K_j$. Therefore, employing the right-hand side of equation (\ref{eq:Sherman-Morrison-Woodbury}) to compute $\bm{\Sigma}^{-1}_{K_j}$ becomes significantly more efficient for large zero-inflated spatial datasets, especially when $n$ is considerably large and $K_j$ is notably smaller than $n$.
 In practice, $K_1$ and $K_2$ are generally much smaller than $n$. This phenomenon is also verified in the simulation study presented in Section \ref{sec5:simulation study}. Therefore, the overall computational complexity of $\bm{\Sigma}^{-1}_{K_1,K_2}$ is reduced to $\mathcal{O}\big(\max \lbrace K_{1}, K_{2} \rbrace^{3} \big)$, instead of $\mathcal{O}(n^{3})$.

 Various basis representations, such as predictive processes (Banerjee et al. 2008), random projections (Guan and Haran 2018; Park and Haran 2020), and multi-resolution basis functions (Nychka et al. 2015; Katzfuss 2017), can be employed for $\{\psi_{j,k}(\cdot): k = 1, \dots, K_j\}$. However, in this paper, we specifically focus on utilizing basis functions derived from TPS.

\subsection{Thin-plate splines}\label{sec3.2:TPS}
 Given data $Z_{1}, \ldots, Z_{n}$ observed at $n$ distinct locations, a TPS function $\psi(\bm{s})$, where $\bm{s} = (s_{1}, s_{2})^t \in \mathcal{R}^{2}$, is determined by minimizing the following penalized least squares criterion:
 \begin{equation}\label{tps obtain}
  \sum\limits_{i = 1}^{n} (Z_{i} - \psi(\bm{s}_{i}))^{2} + \rho\mathcal{J}(\psi),
  \end{equation}
  where
  \begin{equation}\label{smoothness penalty}
  \mathcal{J}(\psi)= \int_{\mathcal{R}^2} \sum_{\upsilon_1+\upsilon_2 = m} \frac{m!}{\upsilon_{1}!\upsilon_{2}!}\Bigg(\frac{\partial^{m}\psi(\bm{s})}{\partial s_{1}^{\upsilon_1}\partial s_{2}^{\upsilon_2}}\Bigg)^{2} d\bm{s}
 \end{equation}
 represents a smoothness penalty, and $\rho \geq 0$ serves as a tuning parameter that controls the smoothness of $\psi(\bm{s})$. For $\rho > 0$ and $m = 2$, the solution to \eqref{tps obtain} falls within the set:
 \begin{align}\label{tps solution}
  \mathcal{F}
  &=
  \Big\lbrace \psi(\cdot):\psi(\bm{s})=\bm{\nu}^{t}\bm{\varphi}(\bm{s})+\varsigma_{0}+\varsigma_{1}s_{1}+\varsigma_{2}s_{2}; \, \bm{\nu} \in \mathcal{R}^{n}, \bm{\varsigma} \in \mathcal{R}^{3}, \bm{\Delta}^{t}\bm{\nu}= \bm{0} \Big\rbrace,
 \end{align}
 where $\bm{\nu} = (\nu_{1},\ldots,\nu_{n})^{t}$, $\bm{\varsigma} = (\varsigma_{0}, \varsigma_{1}, \varsigma_{2})^{t}$,
 \begin{align}\label{Delta}
  \bm{\Delta} =
  \begin{pmatrix}
  1      & s_{1,1} & s_{1,2} \\
  \vdots & \vdots  & \vdots  \\
  1      & s_{n,1} & s_{n,2}
  \end{pmatrix},
 \end{align}
 and
 $\bm{\varphi}(\bm{s})=(\varphi_{1}(\bm{s}),\ldots,\varphi_{n}(\bm{s}))^{t}$ with
 \begin{equation}\label{varphi}
  \varphi_{i}(\bm{s})
  =
  \frac{1}{8\pi} \parallel \bm{s}-\bm{s}_{i} \parallel^{2}\text{log}(\parallel \bm{s}-\bm{s}_{i} \parallel);~~i=1,\dots,n.
 \end{equation}
 The specific form of $\varphi_i(\bm{s})$ is derived from the minimization of a rotation-invariant Sobolev semi-norm in the context of smoothing splines in two-dimensional spaces (Duchon 1977). For further theoretical details, we recommend Duchon (1977) and Wahba (1990), which provide a comprehensive explanation of the mathematical foundations and practical applications of TPS.
 As shown in Theorem 7.1 of Green and Silverman (1993), the penalty term in \eqref{tps obtain} has a clear representation as follows:
 \begin{equation}\label{thm 7.1}
  \mathcal{J}(\psi)=\bm{\nu}^{t}\bm{\Phi}\bm{\nu},
 \end{equation}
 where $\bm{\Phi}$ denotes the $n \times n$ matrix, with the $(i, j)$-th element represented by $\varphi_{j}(\bm{s}_{i})$, specifically:
 \begin{equation}\label{Phi}
  \bm{\Phi}
  =
  \begin{pmatrix}
  \varphi_{1}(\bm{s}_{1}) & \varphi_{2}(\bm{s}_{1}) & \cdots & \varphi_{n}(\bm{s}_{1}) \\
  \vdots & \vdots  & \ddots & \vdots  \\
  \varphi_{1}(\bm{s}_{n}) & \varphi_{2}(\bm{s}_{n}) & \cdots & \varphi_{n}(\bm{s}_{n})
  \end{pmatrix}.
 \end{equation}

 Drawing on the work of Tzeng and Huang (2018), multi-resolution spline basis functions can be derived. Given $\bm{\Delta}$ as specified in \eqref{Delta}, they gave the set of ordered basis functions associated with $\bm{\Delta}$ as follows:
 \begin{equation}\label{tps basis function}
  \psi_{j, k}(\bm{s})=
  \begin{cases}
  1; &\text{if $k=1$,} \\
  s_{k-1}; &\text{if $k=2, 3$,} \\
  \Lambda_{k-3}^{-1}
  \bigg\lbrace \bm{\varphi}(\bm{s})-\bm{\Phi}\bm{\Delta}(\bm{\Delta}^{t}\bm{\Delta})^{-1}\bm{\delta} \bigg\rbrace^{t} \bm{a}_{k-3}; &\text{if $k=4,\ldots,n$,}
  \end{cases}
 \end{equation}
 where $j \in \{1, 2\}$, $\bm{\delta} = (1, \bm{s}^{t})^{t}$, $\bm{\varphi}(\bm{s})$ and $\bm{\Phi}$ are given in \eqref{varphi} and \eqref{Phi}, respectively. Moreover, $\bm{a}_{k}$ is the $k$-th column of $\bm{A}$. The matrix $\bm{A}\:\text{diag}(\Lambda_{1}, \ldots, \Lambda_{n})\bm{A}^{t}$ is the eigen-decomposition of $\bm{Q}\bm{\Phi}\bm{Q}$, with $\Lambda_{1} \geq \cdots \geq \Lambda_{n}$ being the eigenvalues. Here, $\bm{Q} = \bm{I} - \bm{\Delta}(\bm{\Delta}^{t}\bm{\Delta})^{-1}\bm{\Delta}^{t}$, and $\bm{I}$ is an $n \times n$ identity matrix.

\section{Estimation of model parameters}\label{sec4:estimation of model parameters}
 In this section, we detail the methodology for estimating the parameters within the covariance matrix and the regression coefficients as delineated in \eqref{phi} and \eqref{lambda}. Additionally, we describe how the stability of parameter estimation can be assessed using the block jackknife (BJ) method.

\subsection{Parameter estimation for covariance matrices}\label{sec4.1:param for cov matrix}
 The model \eqref{eq:SRE.model} discussed in Section \ref{sec3.1:ApplyTPS} includes parameters $\bm{\Omega}_j$, $\sigma^2_{\xi,j}$, and $K_j$ for $j = 1, 2$, which require estimation. Given the set of basis functions $\{\psi_{j,1}(\bm{s}), \ldots, \psi_{j,K_j}(\bm{s}) : \bm{s} \in \mathcal{D} \text{ and } j = 1, 2\}$, the maximum likelihood (ML) estimates $\widehat{\bm{\Omega}}_j$ and $\hat{\sigma}_{\xi,j}^{2}$ for $\bm{\Omega}_j$ and $\sigma^2_{\xi,j}$ can be derived using the Expectation-Maximization (EM) algorithm. The corresponding closed-form expressions for $\widehat{\bm{\Omega}}_j$ and $\hat{\sigma}_{\xi,j}^{2}$ for $j = 1, 2$, are provided by Tzeng and Huang (2018) as follows:
 \begin{align}
  \widehat{\bm{\Omega}}_{j}
  &=
  \Big( \bm{\Psi}_{j}^{t}\bm{\Psi}_{j} \Big)^{-\frac{1}{2}}\bm{A}_{j}\:\text{diag}\big(\hat{c}_{j, 1}, \ldots, \hat{c}_{j, K_{j}}\big)\bm{A}_{j}^{t}\Big( \bm{\Psi}_{j}^{t}\bm{\Psi}_{j} \Big)^{-\frac{1}{2}}
  \label{MLE for Omega_Kj}
 \end{align}
 and
 \begin{align}
  \hat{\sigma}_{\xi,{j}}^{2}
  &=
  \text{max}\left\{ \frac{1}{n - L^{*}}\Big(\text{tr}(\bm{U}_j) - \sum_{k = 0}^{L^{*}} c_{j, k} \Big), 0 \right\},
  \label{MLE for sigma_xi}
 \end{align}
 where $\bm{U}_j = \bm{Z}_{j}\bm{Z}_{j}^{t}$, $\bm{\Psi}_{j} = (\bm{\psi}_{1}, \ldots, \bm{\psi}_{K_{j}})$, $\bm{\psi}_{k} = (\psi_{j, k}(\bm{s}_{1}), \ldots, \psi_{j, k}(\bm{s}_{n}))^{t}$ for $k = 1, \ldots, K_{j}$, $\bm{A}_{j}\:\text{diag}\big(c_{j, 1}, \ldots, c_{j, K_{j}}\big)\bm{A}_{j}^{t}$ is the eigen-decomposition of $\big( \bm{\Psi}_{j}^{t}\bm{\Psi}_{j} \big)^{-\frac{1}{2}}\bm{\Psi}_{j}^{t}\bm{U}_j\bm{\Psi}_{j}
 \big( \bm{\Psi}_{j}^{t}\bm{\Psi}_{j} \big)^{-\frac{1}{2}}$, $\hat{c}_{j, k} = \text{max}\{c_{j, k} - \hat{\sigma}_{\xi,{j}}^{2}, 0\}, c_{j, 0} = 0$, and $L^{*}$ is the largest $L \in \lbrace1, \ldots, K_{j}\rbrace$ such that $c_{j, L} > \text{max} \lbrace \frac{1}{n - L} (\text{tr}(\bm{U}_j) - $ $\sum_{k = 1}^{L} c_{j, k}), 0 \rbrace$ if it exists, and 0 otherwise. For more detailed information, readers are referred to Theorem 2 in Tzeng and Huang (2018). Here, the usage of the unobserved $\bm{Z}_j$, $j = 1, 2$, is detailed in the last paragraph of Section 4.3.

 The number of basis functions is selected using the extended Akaike’s information criterion (AIC) proposed by Tzeng and Huang (2018), and is determined by $\hat{K}_{j} = \mathop{\arg\min}\limits_{d+1 \leq K_{j} \leq K^{*}} \text{AIC}(K_{j})$, for $j = 1, 2 $, where $K^{*}$ represents an upper bound on the number of basis functions. To balance computational efficiency and modeling flexibility, $K^{*}$ is chosen as 
 $\text{min}\lbrace 10\sqrt{n}, n \rbrace $, as recommended by Tzeng and Huang (2018). This choice ensures a sufficient range of candidate basis functions without incurring excessive computational costs. The detailed derivation and implementation of the AIC criterion, including the corresponding R package \texttt{autoFRK}, are thoroughly described in Tzeng and Huang (2018). This package facilitates efficient computation and practical application of the method in spatial modeling contexts.

 Subsequently, the ML estimates $\widehat{\bm{\Omega}}_{j}$ and $\hat{\sigma}_{\xi,{j}}^{2}$ for $\bm{\Omega}_{j}$ and $\sigma_{\xi,{j}}^{2}$, along with $\hat{K}_j$, are substituted into $\bm{\Sigma}_{K_{j}}$ for $j = 1, 2$. This substitution yields the estimates of the covariance matrices and their inverses, which are displayed as follows:
 \begin{align}
  \widehat{\bm{\Sigma}}_{\hat{K}_{j}}
  &=
  \widehat{\bm{\Psi}}_{j}\widehat{\bm{\Omega}}_{j}\widehat{\bm{\Psi}}_{j}^{t} + \hat{\sigma}_{\xi,{j}}^{2}\bm{I}
  \label{estimate of covariance matrix of Kj}
 \end{align}
 and
 \begin{align}
  \widehat{\bm{\Sigma}}_{\hat{K}_{j}}^{-1}
  &=
  \frac{1}{\hat{\sigma}_{\xi,{j}}^{2}}\bm{I} - \frac{1}{\hat{\sigma}_{\xi,{j}}^{4}}\widehat{\bm{\Psi}}_{j} \Bigg\lbrace \widehat{\bm{\Omega}}_{j}^{-1}+ \frac{1}{\hat{\sigma}_{\xi,{j}}^{2}}\widehat{\bm{\Psi}}_{j}^{t}\widehat{\bm{\Psi}}_{j}  \Bigg\rbrace^{-1} \widehat{\bm{\Psi}}_{j}^{t}.
  \label{inverse of estimate of covariance matrix of Kj}
 \end{align}

 According to equations \eqref{estimate of covariance matrix of Kj} and \eqref{inverse of estimate of covariance matrix of Kj}, we ultimately obtain the estimates of the covariance matrix of $\bm{Z}$ given covariates $\bm{X}$ and its inverse. These estimates are respectively denoted as $\widehat{\bm{\Sigma}}_{\hat{K}_{1}, \hat{K}_{2}}$ and $\widehat{\bm{\Sigma}}_{\hat{K}_{1}, \hat{K}_{2}}^{-1}$, and are provided as follows:
 \begin{align}
  \widehat{\bm{\Sigma}}_{\hat{K}_{1}, \hat{K}_{2}}=
  \Bigg(
  \begin{array}{cc}
  \widehat{\bm{\Sigma}}_{\hat{K}_{1}} & \bm{O} \\
  \bm{O} & \widehat{\bm{\Sigma}}_{\hat{K}_{2}} \\
  \end{array}
  \Bigg)
  \label{estimate of variance of Z given X}
 \end{align}
 and
 \begin{align}
  \widehat{\bm{\Sigma}}_{\hat{K}_{1}, \hat{K}_{2}}^{-1} =
  \Bigg(
  \begin{array}{cc}
  \widehat{\bm{\Sigma}}_{\hat{K}_{1}}^{-1} & \bm{O} \\
  \bm{O} & \widehat{\bm{\Sigma}}_{\hat{K}_{2}}^{-1} \\
  \end{array}
  \Bigg).
  \label{inverse of estimate Z}
 \end{align}

\subsection{Parameter estimation for regression models}\label{sec4.2:param for gee}
 The models \eqref{phi} and \eqref{lambda} discussed in Section \ref{sec2:spatial zero-inflated models} include parameters $\bm{\theta} = (\bm{\beta}^t, \bm{\gamma}^t)^t$, which require estimation. Due to the absence of specific distributional assumptions for the data \( \bm{Y} \), we employ the generalized estimating equations (GEE) approach proposed by Liang and Zeger (1986) to estimate \( \bm{\theta} \). This approach is particularly suitable for spatially correlated zero-inflated count data as it facilitates robust estimation by leveraging the first two moments of the data (specifically, the mean and covariance structure) without requiring the full specification of a likelihood function. Unlike likelihood-based methods, which rely on explicit parametric assumptions about the data distribution (e.g., the Poisson assumption in a ZIP model), the GEE method accommodates potential correlation within the spatial data and provides greater flexibility and robustness in practical applications. This characteristic makes the GEE approach well-suited for scenarios where the true data-generating process may deviate from standard parametric assumptions.

 To introduce the estimation of $\bm{\theta}$, we initially assume that the covariance matrix described in \eqref{eq:Z.covariance.matrix} is known, with $E(Z_{j}(\bm{s}_{i})) = 0$ for $i = 1, \dots, n$ and $j = 1, 2$. Under this assumption, the generalized estimating equation (GEE) for $\bm{\theta}$ is defined as follows, following the approach proposed by He et al. (2015) and Shen and Chen (2024):
 \begin{align}\label{gee}
  U(\bm{\theta}) \equiv \bm{D}^{t}\bm{\Sigma}_{K_{1}, K_{2}}^{-1}\bm{Z} = \bm{0},
 \end{align}
 where $\bm{D}$ is the matrix of partial derivatives, defined as follows:
 \begin{align}
  \bm{D}
  \equiv
  \dfrac{\partial \bm{Z}^{t}}{\partial \bm{\theta}}
  =
  \begin{pmatrix}
  \dfrac{\partial \bm{Z}_{1}}{\partial \bm{\beta}} & \dfrac{\partial \bm{Z}_{2}}{\partial \bm{\beta}} \\
  \dfrac{\partial \bm{Z}_{1}}{\partial \bm{\gamma}} & \dfrac{\partial \bm{Z}_{2}}{\partial \bm{\gamma}} \\
  \end{pmatrix}.
  \label{Matrix D}
 \end{align}
 In \eqref{Matrix D}, the partial derivatives of $\bm{Z}_{1}$ and $\bm{Z}_{2}$ with respect to $\bm{\beta}$ and $\bm{\gamma}$ can be expressed as follows:
 \begin{align}
  \dfrac{\partial Z_{1}(\bm{s}_{i})}{\partial \beta_{l}}
  &=
  -x_{l}(\bm{s}_{i})\bigg(1 - e^{-\lambda(\bm{s}_{i})}\bigg)\phi(\bm{s}_{i})\big(1 - \phi(\bm{s}_{i})\big),
  \label{partial of beta for Z1}
  \\
  \dfrac{\partial Z_{2}(\bm{s}_{i})}{\partial \beta_{l}}
  &=
  0,
  \label{partial of beta for Z2}
  \\
  \dfrac{\partial Z_{1}(\bm{s}_{i})}{\partial \gamma_{l}}
  &=
  x_{l}(\bm{s}_{i})\big(1 - \phi(\bm{s}_{i})\big)\lambda(\bm{s}_{i})e^{-\lambda(\bm{s}_{i})},
  \label{partial of gamma for Z1}
 \end{align}
 and
 \begin{align}
  \dfrac{\partial Z_{2}(\bm{s}_{i})}{\partial \gamma_{l}}
  &=
  -x_{l}(\bm{s}_{i})I(Y(\bm{s}_{i}) > 0)\lambda(\bm{s}_{i})\Big(1 - \big(1 + \lambda(\bm{s}_{i})\big)e^{-\lambda(\bm{s}_{i})}\Big)\Big( 1 - e^{-\lambda(\bm{s}_{i})} \Big)^{-2},
  \label{partial of gamma for Z2}
 \end{align}
 for $l = 0,1,\ldots, p$, and $x_{0}(\bm{s}_{i})=1$.
 Then $\bm{\theta}$ can be estimated by solving equation \eqref{gee}. However, the covariance matrix $\bm{\Sigma}_{K_{1}, K_{2}}$ is typically unknown in practice and must be estimated. As discussed in Section \ref{sec4.1:param for cov matrix}, we can obtain an estimate of the covariance matrix, denoted as $\widehat{\bm{\Sigma}}_{\hat{K}_{1}, \hat{K}_{2}}$. With this estimate, $\widehat{\bm{\Sigma}}_{\hat{K}_{1}, \hat{K}_{2}}$, the iterative procedure for $\bm{\theta}$ at the $(m+1)$-th step can be implemented using the Newton-Raphson method (e.g., McCullagh and Nelder 1989) as follows:
 \begin{equation}\label{newton method for theta}
  \hat{\bm{\theta}}^{(m + 1)}
  =
  \hat{\bm{\theta}}^{(m)} - \bigg(\bm{D}^{(m)^{t}}\widehat{\bm{\Sigma}}_{\hat{K}_{1}, \hat{K}_{2}}^{(m)^{-1}}\bm{D}^{(m)}\bigg)^{-1}\bm{D}^{(m)^{t}}\widehat{\bm{\Sigma}}_{\hat{K}_{1}, \hat{K}_{2}}^{(m)^{-1}}\bm{Z}^{(m)}.
 \end{equation}
 Note that the initial estimates of $\hat{\bm{\theta}}^{(m)}$ (i.e., for $m=0$) can be determined by solving equation (\ref{gee}) using a working independence covariance matrix. Under certain convergence conditions, we can obtain the final estimates $\widehat{\bm{\theta}} = (\hat{\bm{\beta}}^{t}, \hat{\bm{\gamma}}^{t})^{t}$ and $\widehat{\bm{K}} = (\hat{K}_{1}, \hat{K}_{2})$.

 Although the covariance matrix $\bm{\Sigma}_{K_{1}, K_{2}}$ specified in \eqref{gee} may be misspecified, the estimator $\hat{\bm{\theta}}$ still converges due to the consistency properties of GEE, as noted by Liang and Zeger (1986). However, an incorrectly specified covariance structure may result in slow and time-consuming convergence under the GEE framework. To address this, we utilize the TPS technique detailed in Section \ref{sec3.1:ApplyTPS} to capture the underlying covariance structure as accurately as possible without assuming a specific correlation function for the data, thereby accelerating the estimation process.

 Furthermore, since the number of basis functions is substantially smaller than the sample size, computing the inverse of the covariance matrix in a lower-dimensional space becomes feasible and computationally efficient, even with very large sample sizes. This approach overcomes issues related to high-dimensional computations, and its effectiveness is demonstrated in the simulation studies presented in Section \ref{sec5:simulation study}.

\subsection{Estimation algorithm}\label{sec4.3:algorithm}
\noindent
 In this subsection, we comprehensively outline the parameter estimation process for the flexible spatial zero-inflated count model. This process is structured within the GEE framework. The detailed steps of the estimation procedure are organized as follows:
 \begin{enumerate}
    \item \textbf{Input the data:}
    Input the data $\bm{Y} = (Y(\bm{s}_{1}), \ldots, Y(\bm{s}_{n}))^{t}$ into the flexible spatial zero-inflated count model.

    \item \textbf{Construct matrix $\bm{\Phi}$:}
    Construct $\bm{\Delta}$ for $d = 2$ as specified in \eqref{Delta}, and utilize $\bm{\varphi}(\bm{s}) = (\varphi_{1}(\bm{s}), \ldots, \varphi_{n}(\bm{s}))^t$, where each $\varphi_{i}(\bm{s})$ is based on \eqref{varphi}, to assemble the $n \times n$ matrix $\bm{\Phi}$. Each element of $\bm{\Phi}$, specifically the $(i, j)$-th element, is given by $\varphi_{j}(\bm{s}_{i})$.

    \item \textbf{Compute basis functions:}
    Compute the basis functions, $\psi_{j, k}(\bm{s}_{i})$, based on \eqref{tps basis function}.

    \item \textbf{Compute ML estimates:}
    Compute the ML estimates of $\bm{\Omega}_{j}$ and $\sigma_{\xi,{j}}^{2}$ based on \eqref{MLE for Omega_Kj} and \eqref{MLE for sigma_xi}.

    \item \textbf{Determine the number of basis functions using AIC:}
    Use AIC to determine the number of basis functions
    \[
    \hat{K}_{j} = \mathop{\arg\min}\limits_{3 \leq K_{j} \leq K^{*}} \text{AIC}(K_{j})
    \]
    with $K^{*} = \text{min}\lbrace 10\sqrt{n}, n\rbrace$.

    \item \textbf{Compute covariance matrices:}
    Compute $\widehat{\bm{\Sigma}}_{\hat{K}_{1}}$, $\widehat{\bm{\Sigma}}_{\hat{K}_{2}}$, $\widehat{\bm{\Sigma}}_{\hat{K}_{1},\hat{K}_{2}}$ and $\widehat{\bm{\Sigma}}_{\hat{K}_{1},\hat{K}_{2}}^{-1}$.

    \item \textbf{Estimate $\bm{\theta}$:}
    Plug $\widehat{\bm{\Sigma}}_{\hat{K}_{1}, \hat{K}_{2}}^{-1}$ into \eqref{gee} and use \eqref{newton method for theta} to obtain
    \[
    \hat{\bm{\theta}}^{(m+1)} = \left(\hat{\beta}_0^{(m+1)}, \dots, \hat{\beta}_p^{(m+1)}, \hat{\gamma}_0^{(m+1)}, \dots, \hat{\gamma}_p^{(m+1)}\right)^t.
    \]

    \item \textbf{Iterate until convergence:}
    Repeat steps 4 to 7 until
    \[
    \sum_{i=0}^p \left( \left| \hat{\beta}_i^{(m+1)} - \hat{\beta}_i^{(m)} \right| + \left| \hat{\gamma}_i^{(m+1)} - \hat{\gamma}_i^{(m)} \right| \right) < \varepsilon
    \]
    for some prespecified $\varepsilon > 0$.
\end{enumerate}
 \noindent For the computations in Step 4, the values of $\bm{U}_j$ and $\bm{Z}_j$ for $j = 1, 2$ are required. The computation proceeds as follows: First, the estimates of $\hat{\bm{\theta}}$ obtained from equation (\ref{newton method for theta}) are substituted into equations (\ref{phi}) and (\ref{lambda}) to calculate $\phi(\bm{s}_i)$ and $\lambda(\bm{s}_i)$ for $i = 1, \dots, n$. These calculated values are then substituted into equations (\ref{rv for zero(zip-like)}) and (\ref{rv for nonzero(zip-like)}) to derive $Z_1(\bm{s}_i)$ and $Z_2(\bm{s}_i)$. Finally, $\bm{U}_j$ is computed as $\bm{U}_j = \bm{Z}_j \bm{Z}_j^t$.
\subsection{Variance estimation}\label{sec4.4:variance estimation}
 For the GEE-based method, the stability of the estimators is typically evaluated by estimating the variance of $\hat{\bm{\theta}}$ using the sandwich variance estimator (e.g., Hall and Zhang 2004). However, this approach does not perform well in our case because we are working with a single sample of data from $n$ spatial locations. Moreover, the sandwich formula may not provide an accurate approximation due to the inherent spatial correlation in the data. To address this limitation, we adopted the block jackknife (BJ) method, as suggested by Adegboye et al. (2018), which is specifically designed to handle spatially correlated data by dividing the spatial domain into non-overlapping blocks. This approach provides a more robust variance estimation for $\hat{\bm{\theta}}$ by accounting for the spatial dependence within the data. Compared to traditional bootstrapping techniques, the block jackknife method is computationally efficient and avoids potential issues with resampling spatially correlated observations, ensuring more reliable variance estimates.

 In this framework, let $\beta$ denote a specific component of $\bm{\beta}$, and let $\hat{\beta}_{-b}$ represent the estimate of $\beta$ obtained by excluding the $b$th block of data. The BJ method estimates the variance of $\hat{\beta}$ using the following formula:
 \begin{align}\label{block jackknife}
  \widehat{\text{Var}}(\hat{\beta}) \equiv \frac{B - 1}{B} \sum_{b = 1}^{B} (\hat{\beta}_{-b} - \bar{\beta})^{2},
 \end{align}
 where $B$ is the number of blocks, and $\bar{\beta} = B^{-1} \sum_{b = 1}^{B} \hat{\beta}_{-b}$ is the average estimate of $\beta$ across all omitted blocks. A similar approach is used to estimate the variance of each element $\gamma$ in $\bm{\gamma}$, applying the same method as shown in \eqref{block jackknife}. For more details on the block jackknife method, see Sherman (2011, Chapter 10) and Adegboye et al. (2018) for comprehensive discussions on its application and theoretical foundation.

\section{Simulation study}\label{sec5:simulation study}
 In this section, we evaluate the performance of the proposed estimation method for the flexible spatial zero-inflated model, which integrates basis functions with the GEE approach, through numerical simulations. The simulation scenarios encompass various strengths of spatial correlations and different zero-inflation proportions. For parameter estimation, we include additional comparisons by employing a fixed number of basis functions during the regression parameter estimation process. Our numerical studies also encompass scenarios with large datasets to demonstrate the practicality and effectiveness of our method.

\subsection{Setups}\label{sec5.1:setup}
 We define the study region as a unit square $\mathcal{D} = [0, 1] \times [0, 1]$, which is subdivided into $N = 10000 \times 10000 = 10^8$ regular grid points. For each simulation, we randomly select $n = 400$ or $n = 3000$ spatial locations within the study area $\mathcal{D}$. Each selected location is denoted as $\bm{s}_{i} = (s_{i, 1}, s_{i, 2})^{t}$ for $i = 1, \ldots, n$.

 Within the framework of the flexible spatial zero-inflated count models described in Section \ref{sec2:spatial zero-inflated models}, the true mixture probability $\phi(\bm{s}_{i})$ and the true mean rate $\lambda(\bm{s}_{i})$ are specified by the following equations:
 \begin{equation}\label{true phi}
  \text{log}\Bigg(\dfrac{\phi(\bm{s}_{i})}{1 - \phi(\bm{s}_{i})}\Bigg) =\beta_0+\sum_{p=1}^5\beta_p x_p(\bm{s}_i)
 \end{equation}
 and
 \begin{equation}\label{true lambda}
  \text{log}(\lambda(\bm{s}_{i})) =\gamma_0+\sum_{p=1}^5\gamma_p x_p(\bm{s}_i) ,
 \end{equation}
 where $\bm{x}(\bm{s}_{i}) = (1, x_{1}(\bm{s}_{i}), \ldots, x_{5}(\bm{s}_{i}))^{t}$ represents the covariates, and the corresponding regression coefficients are $\bm{\beta} = (\beta_{0}, \beta_{1}, \ldots, \beta_{5})^{t}$ and $\bm{\gamma} = (\gamma_{0}, \gamma_{1}, \ldots, \gamma_{5})^{t}$. The covariates $x_{1}(\bm{s}_{i})$, $x_{2}(\bm{s}_{i})$, and $x_{3}(\bm{s}_{i})$ are independently drawn from a standard normal distribution, while $x_{4}(\bm{s}_{i})$ and $x_{5}(\bm{s}_{i})$ are independently generated from a Bernoulli distribution with a success probability of 0.5. The values of $\bm{\beta}$ and $\bm{\gamma}$ are chosen to ensure that the zero-inflation proportions of the count responses are maintained at either 40\% or 70\%.

 To generate the zero-inflated count responses $\lbrace Y(\bm{s}_{i}): i = 1, \ldots, n \rbrace$, we follow the approach outlined by Shen and Chen (2024) by first generating a set of correlated binary variables $\lbrace O(\bm{s}_{i}) : i = 1, \ldots, n \rbrace$. These binary variables determine whether each response in $\lbrace Y(\bm{s}_{i}): i = 1, \ldots, n \rbrace$ is a zero or is drawn from a Poisson distribution. Specifically, we start by generating $n$ correlated random variables from an $n$-dimensional normal distribution with zero mean, unit variance, and a correlation matrix defined by the exponential correlation model (e.g., Cressie 1993):
 \begin{equation}\label{exponential correlation}
  \rho_{j}(h_{m, n}; \sigma_{j}^{2}, \tau_{j}^{2}, \delta_{j}) =
  \text{Corr}(Z_{j}(\bm{s}_{m}), Z_{j}(\bm{s}_{n}))
  =\dfrac{\sigma_{j}^{2}}{\sigma_{j}^{2} + \tau_{j}^{2}}\text{exp}(-h_{m, n}/\delta_{j});\ j = 1, 2,
 \end{equation}
 where the sill parameter $\sigma_{j}^{2}$ is set to 1, the nugget parameter $\tau_{j}^{2}$ is set to 2, and the range parameter $\delta_{j}$ is set to $0.3 \times h_{max}$, with $h_{m, n} = ~\parallel\bm{s}_{m}-\bm{s}_{n}\parallel$ and $h_{max} = \sqrt{2}$ representing the maximum distance within the study region $\mathcal{D} = [0, 1]^2$.
 These $n$ normal random variables are then transformed into $n$ correlated binary variables $\lbrace O(\bm{s}_{i}) : i = 1, \ldots, n \rbrace$ via the quantile-probability transformation. Following a similar procedure, but adjusting the parameter $\delta_{j}$ to $c \times h_{max}$, we generate $n$ correlated Poisson variables, where a larger value of $c$ indicates stronger spatial correlation among the responses. Finally, to ensure that the Poisson variables align with the marginal means specified in \eqref{true lambda}, we assign $Y(\bm{s}_{i}) = 0$ if $O(\bm{s}_{i}) = 1$, indicating that the observation corresponds to a zero count. Otherwise, if $O(\bm{s}_{i}) = 0$, $Y(\bm{s}_{i})$ is drawn from the corresponding Poisson distribution. 
 
 It is important to clarify that this data-generating mechanism does not correspond to a standard ZIP model. While the non-zero counts are conditionally sampled from a Poisson distribution, the overall process combines a spatially correlated binary component and spatially varying Poisson rates. This construction leads to a joint distribution of $Y(\bm{s})$ that is not analytically tractable and does not satisfy the full likelihood structure assumed in classical ZIP models. Moreover, the binary indicator process is not independent of the count-generating process, and the marginal distribution of $Y(\bm{s})$ includes zero inflation in a nonparametric and spatially heterogeneous manner. This breaks the separability assumption of classical ZIP models where structural zeros and count components are often modeled independently. Therefore, although a Poisson distribution is used as a building block, the resulting data do not conform to any closed-form ZIP or Poisson model. Instead, they satisfy only the first-moment conditions of our proposed model (Equations~\eqref{zero mean} and~\eqref{nonzero mean}), and do not meet the stronger assumptions required by fully parametric ZIP or Poisson likelihoods. This setting allows us to evaluate the robustness of the proposed estimation method under distributional misspecification, which is commonly encountered in practical spatial count data.

 In addition to evaluating spatial covariance structures determined by a fixed number of basis functions ($K_{1} = K_{2} = 30$) or selected using AIC, we also investigate scenarios with varying intensities of spatial correlation (Low: $c = 0.01$; Medium: $c = 0.3$; High: $c = 0.8$) and different levels of zero inflation in the responses (40\% and 70\%). These scenarios are tested with both small ($n = 400$) and large ($n = 3000$) sample sizes to demonstrate the effectiveness of the proposed iterative estimation algorithm. The results from 200 simulations, using $\varepsilon = 0.03$ as the convergence criterion in Step 8 of the algorithm, are summarized in Tables \ref{tab:small sample estimation(40 percent)}, \ref{tab:small sample estimation(70 percent)}, and \ref{tab:large sample estimation(40 percent)}. In each table, the first column presents the true values of $\bm{\theta} = (\bm{\beta}^{t}, \bm{\gamma}^{t})^{t}$ across the different scenarios. 

 To avoid possible misinterpretation, it is important to note that the notation ``AIC'' reported in the following Tables solely indicates the number of basis functions determined by the AIC-based selection procedure (Tzeng and Huang 2018) rather than likelihood-based model comparisons.
 
\subsection{Results}\label{sec5.2:results of parameter}
\subsubsection{Small samples}\label{sec5.2.1:small samples}
 In Tables \ref{tab:small sample estimation(40 percent)} and \ref{tab:small sample estimation(70 percent)}, $ \hat{\bm{\theta}}_{\text{AIC}}$ and $\hat{\bm{\theta}}_{\text{Fix}}$ represent the parameter estimates obtained from the proposed iterative estimation algorithm. Specifically, $\hat{\bm{\theta}}_{\text{AIC}}$ is derived using the number of basis functions selected by Akaike’s information criterion (AIC), while $\hat{\bm{\theta}}_{\text{Fix}}$ is based on a fixed number of basis functions. Owing to the consistency properties of the GEE framework, both $\hat{\bm{\theta}}_{\text{AIC}}$ and $\hat{\bm{\theta}}_{\text{Fix}}$ consistently yield accurate estimates across all scenarios. This demonstrates superior robustness and flexibility of the proposed approach, particularly in situations where the true underlying distribution is either unknown or misspecified.
 
 For further analysis of $\hat{\bm{\theta}}_{\text{AIC}}$ and $\hat{\bm{\theta}}_{\text{Fix}}$, we estimate their variances using the block jackknife (BJ) method as described in Section \ref{sec4.4:variance estimation}. In applying the BJ method, the dataset is divided into $B = 20$ blocks via the k-means clustering algorithm. We then construct approximate 95\% confidence intervals for each simulation to calculate the coverage probabilities of $\bm{\theta}$. These coverage probabilities, based on 200 simulations, are presented in Table \ref{tab:small sample CP}.
 The results indicate that $\hat{\bm{\theta}}_{\text{AIC}}$ generally performs better than $\hat{\bm{\theta}}_{\text{Fix}}$, requiring fewer iterations as shown in Table \ref{tab:the number of basis functions and iteration for small sample}. This improvement is due to the optimal selection of basis functions by AIC, which better approximates the underlying spatial structure. Furthermore, Table \ref{tab:the number of basis functions and iteration for small sample} lists the average number of basis functions selected by AIC across different scenarios, demonstrating that a smaller number of basis functions significantly reduces the dimensionality challenges associated with computing the inverse matrix, thereby enhancing the efficiency of parameter estimation in the proposed method.
 
 Likelihood-based methods, such as the spatial ZIP model, rely on strong distributional assumptions to model the entire data structure. In contrast, our proposed GEE-based model only requires assumptions about the means of the responses (e.g., Equations \eqref{zero mean} and \eqref{nonzero mean}) without relying on full distributional assumptions. This distinction makes the GEE-based approach more flexible and robust in practical applications, particularly when the true data-generating process deviates from the assumed distribution. The findings of Shen and Chen (2024) (e.g., the spatial ZIP model) have demonstrated through simulation studies that likelihood-based methods are sensitive to the underlying data structure. By comparison, the GEE-based approach, which relies on moment-based inference, tends to exhibit greater robustness under various data scenarios. Readers interested in further details about the spatial ZIP model are encouraged to consult Shen and Chen (2024). In light of these findings, we build upon the foundation established by Shen and Chen (2024) and focus on extending the GEE framework by incorporating a more comprehensive modeling of the covariance structure, thereby enhancing its applicability to large-scale spatial zero-inflated count data.
 
 To assess the accuracy in estimating the latent spatial structures, we evaluated the mean squared errors (MSEs) of the proposed estimators for $\lambda(\bm{s})$ and $\phi(\bm{s})$ under various scenarios. These MSEs were computed based on 200 replicates and compared the performance of two approaches: the ZIP generalized additive model (GAM) with linear covariates and a smooth spatial effect (e.g., Wood 2017), and our proposed GEE-based method. The analysis was conducted using simulated datasets with $n=400$ spatial locations. The results (see Table \ref{tab:prediction}) indicate that our GEE-based method consistently achieves lower MSEs compared to the GAM approach, demonstrating its superior ability to recover the true spatial surfaces. This improvement can be attributed to the flexibility and robustness of the TPS-based basis expansion in capturing spatial variability while effectively leveraging the GEE framework to account for spatial dependence. These findings are consistent with the theoretical advantages of moment-based methods, such as GEE, which avoid restrictive distributional assumptions while focusing on marginal mean and covariance structures (Liang and Zeger 1986). By contrast, ZIP-GAM, which relies on likelihood-based estimation, may be sensitive to the assumed distributional structure of the data, potentially leading to reduced prediction accuracy in zero-inflated count data scenarios. This result highlights the strength of the proposed method in practical applications where accurate spatial prediction is critical.

 Table \ref{tab:average time for small sample} presents the average computation time required for parameter estimation across 200 replicates under various scenarios. The results indicate that using a fixed number of basis functions for constructing the spatial structure is significantly faster than the approach that involves selecting basis functions via AIC. This difference in time is primarily due to the AIC selection process, which involves evaluating a larger set of basis functions (e.g., $\text{min}\lbrace 10\sqrt{n}, n\rbrace$) before finalizing the selection, leading to increased computation time. Nevertheless, the overall estimation time remains relatively brief, ensuring that the proposed iterative estimation procedure efficiently achieves optimal results. This underscores the efficiency of the proposed method.

\begin{table}[!h]
\caption{Means and standard deviations of parameter estimation results for $n = 400$ with a 40\% zero-inflation proportion under varying spatial correlation strengths (Low: $c = 0.01$; Medium: $c = 0.3$; High: $c = 0.8$) based on 200 replicates.}
\begin{center}
 \footnotesize
 \begin{tabular}{ccrrrrrrr}
 \toprule
 &\multicolumn{2}{c}{$\bm{\theta}$}&\multicolumn{2}{c}{$\hat{\bm{\theta}}_{\rm AIC}$}&\multicolumn{2}{c}{$\hat{\bm{\theta}}_{\rm Fix}$}\\
 \cmidrule(rl){2-3}\cmidrule(rl){4-5}\cmidrule(rl){6-7}
 Correlation&&True&mean&s.d.&mean&s.d.\\
 \midrule
 Low& $\beta_{0}$   & -0.7 & -0.680 & 0.395  & -0.695 & 0.456   \\
    & $\beta_{1}$   & -0.6 & -0.690 & 0.261  & -0.684 & 0.243   \\
    & $\beta_{2}$   & -0.6 & -0.716 & 0.250  & -0.730 & 0.266   \\
    & $\beta_{3}$   & -0.6 & -0.694 & 0.256  & -0.696 & 0.263   \\
    & $\beta_{4}$   & -0.5 & -0.578 & 0.530  & -0.564 & 0.555   \\
    & $\beta_{5}$   & -0.5 & -0.567 & 0.395  & -0.521 & 0.400   \\
    & $\gamma_{0}$  &  0.4 &  0.405 & 0.130  &  0.394 & 0.140   \\
    & $\gamma_{1}$  &  0.3 &  0.298 & 0.055  &  0.298 & 0.054  \\
    & $\gamma_{2}$  &  0.3 &  0.297 & 0.061  &  0.299 & 0.061  \\
    & $\gamma_{3}$  &  0.3 &  0.297 & 0.059  &  0.299 & 0.061  \\
    & $\gamma_{4}$  & -0.3 & -0.308 & 0.107  & -0.311 & 0.108  \\
    & $\gamma_{5}$  &  0.6 &  0.593 & 0.113  &  0.600 & 0.116  \\
 \bottomrule
 Medium & $\beta_{0}$   & -0.7 & -0.708 & 0.389 & -0.705 & 0.457  \\
        & $\beta_{1}$   & -0.6 & -0.693 & 0.256 & -0.689 & 0.254  \\
        & $\beta_{2}$   & -0.6 & -0.716 & 0.260 & -0.747 & 0.275  \\
        & $\beta_{3}$   & -0.6 & -0.698 & 0.250 & -0.703 & 0.249  \\
        & $\beta_{4}$   & -0.5 & -0.687 & 0.513 & -0.681 & 0.557  \\
        & $\beta_{5}$   & -0.5 & -0.496 & 0.419 & -0.488 & 0.448  \\
        & $\gamma_{0}$  &  0.4 &  0.355 & 0.162 &  0.334 & 0.175  \\
        & $\gamma_{1}$  &  0.3 &  0.304 & 0.054 &  0.306 & 0.056 \\
        & $\gamma_{2}$  &  0.3 &  0.305 & 0.060 &  0.305 & 0.060 \\
        & $\gamma_{3}$  &  0.3 &  0.304 & 0.060 &  0.307 & 0.060  \\
        & $\gamma_{4}$  & -0.3 & -0.323 & 0.106 & -0.328 & 0.102  \\
        & $\gamma_{5}$  &  0.6 &  0.618 & 0.108 &  0.630 & 0.112  \\
 \bottomrule
 High & $\beta_{0}$   & -0.7 & -0.697 & 0.348 & -0.686 & 0.384  \\
      & $\beta_{1}$   & -0.6 & -0.675 & 0.242 & -0.692 & 0.258  \\
      & $\beta_{2}$   & -0.6 & -0.709 & 0.250 & -0.742 & 0.263  \\
      & $\beta_{3}$   & -0.6 & -0.705 & 0.261 & -0.709 & 0.267  \\
      & $\beta_{4}$   & -0.5 & -0.694 & 0.537 & -0.695 & 0.544  \\
      & $\beta_{5}$   & -0.5 & -0.518 & 0.404 & -0.512 & 0.414  \\
      & $\gamma_{0}$  &  0.4 &  0.341 & 0.149 &  0.335 & 0.160  \\
      & $\gamma_{1}$  &  0.3 &  0.310 & 0.051 &  0.309 & 0.053  \\
      & $\gamma_{2}$  &  0.3 &  0.307 & 0.057 &  0.306 & 0.059  \\
      & $\gamma_{3}$  &  0.3 &  0.305 & 0.055 &  0.308 & 0.057  \\
      & $\gamma_{4}$  & -0.3 & -0.324 & 0.115 & -0.333 & 0.108  \\
      & $\gamma_{5}$  &  0.6 &  0.615 & 0.098 &  0.622 & 0.100  \\
 \bottomrule
 \label{tab:small sample estimation(40 percent)}
 \end{tabular}
 \end{center}
\end{table} 

\begin{table}[!h]
\caption{Means and standard deviations of parameter estimation results for $n = 400$ with a 70\% zero-inflation proportion under varying spatial correlation strengths (Low: $c = 0.01$; Medium: $c = 0.3$; High: $c = 0.8$) based on 200 replicates.}
\begin{center}
 \footnotesize
 \begin{tabular}{ccrrrrrrr}
 \toprule
 &\multicolumn{2}{c}{$\bm{\theta}$}&\multicolumn{2}{c}{$\hat{\bm{\theta}}_{\rm AIC}$}&\multicolumn{2}{c}{$\hat{\bm{\theta}}_{\rm Fix}$}\\
 \cmidrule(rl){2-3}\cmidrule(rl){4-5}\cmidrule(rl){6-7}
 Correlation&&True&mean&s.d.&mean&s.d.\\
 \midrule
 Low& $\beta_{0}$   & 0.5  &  0.454 & 0.498 &  0.474 & 0.549  \\
    & $\beta_{1}$   & 0.6  &  0.641 & 0.162 &  0.647 & 0.164  \\
    & $\beta_{2}$   & 0.5  &  0.594 & 0.231 &  0.599 & 0.248  \\
    & $\beta_{3}$   & 0.5  &  0.568 & 0.212 &  0.575 & 0.224  \\
    & $\beta_{4}$   & 0.5  &  0.538 & 0.368 &  0.543 & 0.397  \\
    & $\beta_{5}$   & -0.5 & -0.510 & 0.392 & -0.511 & 0.399  \\
    & $\gamma_{0}$  & 0.3  &  0.238 & 0.210 &  0.231 & 0.227  \\
    & $\gamma_{1}$  & -0.3 & -0.313 & 0.084 & -0.315 & 0.086 \\
    & $\gamma_{2}$  & 0.5  &  0.518 & 0.086 &  0.518 & 0.088 \\
    & $\gamma_{3}$  & -0.5 & -0.508 & 0.093 & -0.509 & 0.095  \\
    & $\gamma_{4}$  & -0.6 & -0.636 & 0.183 & -0.634 & 0.190  \\
    & $\gamma_{5}$  & 0.6  &  0.630 & 0.174 &  0.635 & 0.185  \\
 \bottomrule
 Medium & $\beta_{0}$   & 0.5  &  0.421 & 0.469 &  0.421 & 0.508  \\
        & $\beta_{1}$   & 0.6  &  0.647 & 0.163 &  0.641 & 0.168  \\
        & $\beta_{2}$   & 0.5  &  0.636 & 0.245 &  0.637 & 0.249  \\
        & $\beta_{3}$   & 0.5  &  0.565 & 0.216 &  0.564 & 0.207  \\
        & $\beta_{4}$   & 0.5  &  0.525 & 0.384 &  0.512 & 0.397  \\
        & $\beta_{5}$   & -0.5 & -0.492 & 0.364 & -0.487 & 0.366  \\
        & $\gamma_{0}$  & 0.3  &  0.215 & 0.266 &  0.215 & 0.248  \\
        & $\gamma_{1}$  & -0.3 & -0.315 & 0.087 & -0.316 & 0.084  \\
        & $\gamma_{2}$  & 0.5  &  0.528 & 0.093 &  0.527 & 0.092  \\
        & $\gamma_{3}$  & -0.5 & -0.513 & 0.099 & -0.514 & 0.091  \\
        & $\gamma_{4}$  & -0.6 & -0.631 & 0.194 & -0.634 & 0.181  \\
        & $\gamma_{5}$  & 0.6  &  0.637 & 0.190 &  0.635 & 0.181  \\
 \bottomrule
 High & $\beta_{0}$   & 0.5  &  0.405 & 0.476 &  0.417 & 0.526  \\
      & $\beta_{1}$   & 0.6  &  0.667 & 0.181 &  0.676 & 0.189  \\
      & $\beta_{2}$   & 0.5  &  0.646 & 0.244 &  0.651 & 0.259  \\
      & $\beta_{3}$   & 0.5  &  0.578 & 0.228 &  0.585 & 0.235  \\
      & $\beta_{4}$   & 0.5  &  0.530 & 0.391 &  0.536 & 0.419  \\
      & $\beta_{5}$   & -0.5 & -0.502 & 0.363 & -0.503 & 0.376  \\
      & $\gamma_{0}$  & 0.3  &  0.202 & 0.266 &  0.200 & 0.264  \\
      & $\gamma_{1}$  & -0.3 & -0.313 & 0.087 & -0.314 & 0.087  \\
      & $\gamma_{2}$  & 0.5  &  0.532 & 0.090 &  0.531 & 0.091  \\
      & $\gamma_{3}$  & -0.5 & -0.517 & 0.096 & -0.517 & 0.092  \\
      & $\gamma_{4}$  & -0.6 & -0.642 & 0.184 & -0.640 & 0.182  \\
      & $\gamma_{5}$  & 0.6  &  0.643 & 0.187 &  0.641 & 0.182  \\
 \bottomrule
 \label{tab:small sample estimation(70 percent)}
 \end{tabular}
 \end{center}
\end{table} 

\begin{table}[!h]
\caption{Coverage probabilities of each parameter under different zero-inflation proportions and spatial correlation strengths (Low: $c = 0.01$; Medium: $c = 0.3$; High: $c = 0.8$) at $n = 400$ based on 200 replicates. ``AIC'' denotes the number of basis functions selected using Akaike's information criterion, while ``Fix'' indicates a fixed number of basis functions.}
\begin{center}
 \footnotesize
 \begin{tabular}{cccccc}
 \toprule
 &&\multicolumn{2}{c}{40\%}&\multicolumn{2}{c}{70\%}\\
 \cmidrule(rl){3-4}\cmidrule(rl){5-6}
 Correlation&Parameter&AIC&Fix&AIC&Fix\\
 \midrule
 Low& $\beta_{0}$     &  0.980 & 0.976 &  0.941 & 0.912  \\
    & $\beta_{1}$     &  0.995 & 0.995 &  0.995 & 0.990  \\
    & $\beta_{2}$     &  0.976 & 0.967 &  0.921 & 0.900  \\
    & $\beta_{3}$     &  0.990 & 0.986 &  0.983 & 0.975  \\
    & $\beta_{4}$     &  0.928 & 0.912 &  0.971 & 0.945  \\
    & $\beta_{5}$     &  0.985 & 0.976 &  0.954 & 0.954  \\
    & $\gamma_{0}$    &  0.947 & 0.945 &  0.962 & 0.958  \\
    & $\gamma_{1}$    &  0.971 & 0.963 &  0.941 & 0.929  \\
    & $\gamma_{2}$    &  0.910 & 0.903 &  0.962 & 0.962  \\
    & $\gamma_{3}$    &  0.900 & 0.901 &  0.929 & 0.929  \\
    & $\gamma_{4}$    &  0.952 & 0.945 &  0.971 & 0.962  \\
    & $\gamma_{5}$    &  0.942 & 0.958 &  0.954 & 0.954  \\
 \bottomrule
 Medium & $\beta_{0}$     &  0.989 & 0.982 &  0.950 & 0.944  \\
        & $\beta_{1}$     &  0.976 & 0.976 &  0.992 & 0.991  \\
        & $\beta_{2}$     &  0.982 & 0.964 &  0.900 & 0.901  \\
        & $\beta_{3}$     &  0.988 & 0.992 &  0.979 & 0.978  \\
        & $\beta_{4}$     &  0.923 & 0.917 &  0.958 & 0.948  \\
        & $\beta_{5}$     &  0.988 & 0.958 &  0.963 & 0.953  \\
        & $\gamma_{0}$    &  0.912 & 0.917 &  0.934 & 0.948  \\
        & $\gamma_{1}$    &  0.941 & 0.934 &  0.954 & 0.944  \\
        & $\gamma_{2}$    &  0.953 & 0.958 &  0.934 & 0.944  \\
        & $\gamma_{3}$    &  0.929 & 0.929 &  0.950 & 0.940  \\
        & $\gamma_{4}$    &  0.935 & 0.947 &  0.946 & 0.965  \\
        & $\gamma_{5}$    &  0.964 & 0.953 &  0.954 & 0.953  \\
 \bottomrule
 High & $\beta_{0}$     &  0.986 & 0.980 &  0.949 & 0.930  \\
      & $\beta_{1}$     &  0.979 & 0.980 &  0.988 & 0.980  \\
      & $\beta_{2}$     &  0.959 & 0.961 &  0.901 & 0.900  \\
      & $\beta_{3}$     &  0.986 & 0.986 &  0.976 & 0.957  \\
      & $\beta_{4}$     &  0.900 & 0.902 &  0.953 & 0.949  \\
      & $\beta_{5}$     &  0.994 & 0.993 &  0.961 & 0.949  \\
      & $\gamma_{0}$    &  0.901 & 0.922 &  0.907 & 0.918  \\
      & $\gamma_{1}$    &  0.932 & 0.928 &  0.934 & 0.934  \\
      & $\gamma_{2}$    &  0.926 & 0.941 &  0.930 & 0.926  \\
      & $\gamma_{3}$    &  0.926 & 0.915 &  0.942 & 0.945  \\
      & $\gamma_{4}$    &  0.926 & 0.935 &  0.969 & 0.968  \\
      & $\gamma_{5}$    &  0.966 & 0.973 &  0.942 & 0.953  \\
 \bottomrule
 \label{tab:small sample CP}
 \end{tabular}
 \end{center}
\end{table} 

\begin{table}[htbp] 
\caption{Average number of basis functions and iterations for different zero-inflation proportions and spatial correlations at $n = 400$ across 200 replicates. Standard deviations are provided in parentheses. ``AIC'' represents the number of basis functions selected by Akaike's information criterion, while ``Fix'' indicates a fixed number of basis functions.}

\begin{center}
\begin{tabular}{ccccccc} \toprule
& \multicolumn{2}{c}{$\hat{K}_{1}$} & \multicolumn{2}{c}{$\hat{K}_{2}$} & \multicolumn{2}{c}{number of iterations}
\\
\cmidrule(r){2-5} \cmidrule(r){6-7}
Method& 40\% & 70\% & 40\% & 70\% & 40\% & 70\%
\\
\cline{2-7}
\multicolumn{7}{c}{Low correlation: $c = 0.01$}
\\
\hline
\multirow{2}*{AIC} & 6.00 & 7.00 & 4.00 & 9.00 & 8.81 & 7.25 \\
			~ & (0.42) & (0.43) & (0.40) & (1.18) & (4.44) & (1.77) \\
\multirow{2}*{Fix} & 30.00 & 30.00 & 30.00 & 30.00 & 11.88 & 8.01 \\
			~ & (0.00) & (0.00) & (0.00) & (0.00) & (4.83) & (2.05) \\
\cline{2-7}
\multicolumn{7}{c}{Moderate correlation: $c = 0.3$}
\\
\hline
\multirow{2}*{AIC} & 8.00 & 9.00 & 9.00 & 10.00 & 9.52 & 7.31 \\
			~ & (0.56) & (0.58) & (0.70) & (1.35) & (3.46) & (1.66) \\
\multirow{2}*{Fix} & 30.00 & 30.00 & 30.00 & 30.00 & 11.46 & 7.95 \\
			~ & (0.00) & (0.00) & (0.00) & (0.00) & (4.02) & (1.95)
\\
\cline{2-7}
\multicolumn{7}{c}{High correlation: $c = 0.8$}
\\
\hline
\multirow{2}*{AIC} & 7.00 & 8.00 & 7.00 & 12.00 & 8.86 & 7.35 \\
			~ & (0.53) & (0.49) & (0.67) & (1.54) & (2.64) & (1.78) \\
\multirow{2}*{Fix} & 30.00 & 30.00 & 30.00 & 30.00 & 11.14 & 7.94 \\
			~ & (0.00) & (0.00) & (0.00) & (0.00) & (3.14) & (1.96)
\\
\bottomrule
\end{tabular}
\end{center}
\label{tab:the number of basis functions and iteration for small sample}
\end{table} 

\begin{table}[htbp]
\caption{Mean squared errors (MSEs) of the estimated $\phi(\bm{s})$ and $\lambda(\bm{s})$ under various scenarios, comparing results obtained using GAM and GEE methods based on $n=400$ and 200 replicates. Values in parentheses represent the corresponding standard errors.}
\begin{center}
        \begin{tabular}{lcccccccc}\toprule
        &\multicolumn{3}{c}{40\%}  &&\multicolumn{3}{c}{70\%}    \\ 
        \cline{2-4}                \cline{6-8}                   
        &$c=0.01$& $c=0.3$& $c=0.8$&&$c=0.01$& $c=0.3$& $c=0.8$  \\
        \cline{2-4}                \cline{6-8}                              
        $\hat{\phi}_{\text{GAM}}$     & 0.0282 & 0.0318 & 0.0933 && 0.0233 & 0.0294 & 0.0286  \\
                                        &(0.0005)&(0.0009)&(0.0021)&&(0.0059)&(0.0096)& (0.0100)\\
        $\hat{\phi}_{\text{GEE}}$       & 0.0122 & 0.0139 & 0.0141 && 0.0121 & 0.0173 & 0.0166  \\
                                        &(0.0008)&(0.0007)&(0.0007)&&(0.0008)&(0.0014)& (0.0011)\\ 
                                        \cline{2-4}                \cline{6-8}  
        $\hat{\lambda}_{\text{GAM}}$  & 0.1064 & 0.2488 & 0.2287 && 0.2836 & 0.4424 & 0.4448  \\
                                        &(0.0039)&(0.0116)&(0.0103)&&(0.0116)&(0.0225)& (0.0301)\\
        $\hat{\lambda}_{\text{GEE}}$    & 0.1029 & 0.1291 & 0.1324 && 0.2865 & 0.3219 & 0.3245  \\ 
                                        &(0.0042)&(0.0061)&(0.0070)&&(0.0145)&(0.0181)& (0.0208)\\
        \bottomrule
        \label{tab:prediction}
        \end{tabular}
        \end{center}
    \end{table} 
 
\begin{table}[htbp] 
\caption{Average time required per replicate for different zero-inflation proportions and spatial correlations at $n = 400$. ``AIC'' represents the number of basis functions selected by Akaike's information criterion, while ``Fix'' indicates a fixed number of basis functions.}
\begin{center}
\begin{tabular}{cccc} \toprule
\multicolumn{4}{c}{40\% zero-inflated}
\\
\cline{2-4}
 & Low correlation & Moderate correlation & High correlation
\\
\hline
AIC & 9.76 sec & 13.38 sec & 14.24 sec \\
Fix & 7.38 sec & 4.97 sec & 4.89 sec \\
\hline
\multicolumn{4}{c}{70\% zero-inflated}
\\
\cline{2-4}
 & Low correlation & Moderate correlation & High correlation
\\
\hline
AIC & 8.66 sec & 9.33 sec & 9.19 sec \\
Fix & 4.66 sec & 4.81 sec & 4.96 sec \\
\bottomrule
\end{tabular}
\end{center}
\label{tab:average time for small sample}
\end{table} 

\subsubsection{Large samples}\label{sec5.2.2:large samples}
\noindent
 In this subsection, the settings are largely consistent with those used in previous cases. We analyze a large sample size of $n=3000$ with a 40\% zero-inflation proportion and a moderate spatial correlation (i.e., $c=0.3$) to illustrate the efficiency of the proposed iterative estimation procedure in parameter estimation. The notation for parameter estimation remains unchanged from that in the small sample size scenarios. Table \ref{tab:large sample estimation(40 percent)} presents the parameter estimation results, showing that both $\hat{\bm{\theta}}_{\text{AIC}}$ and $\hat{\bm{\theta}}_{\text{Fix}}$ perform similarly, with estimates closely aligning with the true values $\bm{\theta}$, consistent with the properties of the GEE framework.

 As in the small sample scenario, we use the k-means clustering algorithm to partition the data into 20 blocks and apply the BJ method to estimate the variances of the parameter estimators. Approximate 95\% confidence intervals for $\bm{\theta} = (\bm{\beta}^{t}, \bm{\gamma}^{t})^{t}$ are then constructed, and the corresponding coverage probabilities (CP) are calculated. Both $\hat{\bm{\theta}}_{\text{AIC}}$ and $\hat{\bm{\theta}}_{\text{Fix}}$ generally exhibit similar and satisfactory performance in terms of coverage probabilities. For the few instances (i.e., $\beta_4$ and $\gamma_0$ in Table \ref{tab:large sample estimation(40 percent)}) where coverage probabilities fell below expectations, this may be attributed to an insufficient number of iterations or a convergence criterion that was not stringent enough. These factors could lead to slight biases in the parameter estimates, which in turn affect the coverage probabilities. This finding underscores the importance of carefully setting and evaluating the convergence criteria in iterative estimation procedures. Determining appropriate convergence criteria is a critical step in such procedures, but it also presents a significant challenge, especially when multiple parameters are estimated simultaneously. Achieving perfect convergence for all parameters can be particularly demanding. Nevertheless, under the convergence criteria considered in this study (see Step 8 of the estimation algorithm in Section 4.3), the results presented here represent a notable advancement and are satisfactory for analyzing spatially correlated zero-inflated count data using the proposed GEE-based model.

 In Table \ref{tab:number of basis functions and iteration for large sample(40 percent)}, as with the small sample scenario, we report the average number of basis functions and iterations across 200 replicates. The table shows that with no more than 30 basis functions, it is possible to construct a $3000 \times 3000$ matrix, making the computation of its inverse feasible and effectively addressing the challenges of high-dimensional matrix inversion. Additionally, the data in Table \ref{tab:number of basis functions and iteration for large sample(40 percent)} reveal that under the specified convergence criterion (i.e., $\varepsilon=0.03$ in Step 8 of the iterative estimation algorithm), approximately six iterations are sufficient to obtain regression parameter estimates, demonstrating that our method remains efficient even with large datasets.

 Finally, we documented the average time required for parameter estimation in each replicate. The results revealed a trend similar to that observed in small sample scenarios concerning the selection of basis functions, whether determined by AIC or fixed in advance. Specifically, constructing the spatial structure with a fixed number of basis functions required an average of 34.8 seconds per replicate, whereas using AIC to select the basis functions took an average of 49.2 seconds. Nevertheless, the parameter estimation time for both methods averaged less than one minute, achieving convergence and final parameter estimates efficiently. This demonstrates that the proposed iterative algorithm is not only suitable for large datasets but also effective in execution.

\begin{table}[htbp] 
\caption{Parameter estimation results for $n = 3000$ with a $40\%$ zero-inflation proportion and a moderate spatial correlation ($c = 0.3$) using the proposed iterative estimation algorithm with 200 replicates. ``BJ'' represents the block jackknife method, while ``CP'' indicates the coverage probability.}
\begin{center}
\begin{tabular}{c  r r  c r  c c  c c  c}
\toprule 
\multicolumn{2}{c}{$\bm{\theta}$}&\multicolumn{2}{c}{$\hat{\bm{\theta}}_{\rm AIC}$}&\multicolumn{2}{c}{$\hat{\bm{\theta}}_{\rm Fix}$}&\multicolumn{2}{c}{BJ}&\multicolumn{2}{c}{CP}\\
 \cmidrule(rl){1-2}\cmidrule(rl){3-4}\cmidrule(rl){5-6}\cmidrule(rl){7-8}\cmidrule(rl){9-10}
 &True&mean&s.d.&mean&s.d.&$\hat{\bm{\theta}}_{\rm AIC}$&$\hat{\bm{\theta}}_{\rm Fix}$&$\hat{\bm{\theta}}_{\rm AIC}$&$\hat{\bm{\theta}}_{\rm Fix}$\\
\midrule
$\beta_{0}$ & -0.7 & -0.793 & 0.119 & -0.796 & 0.119 & 0.191 & 0.190 & 0.993 & 0.992
\\
$\beta_{1}$ & -0.6 & -0.708 & 0.078 & -0.705 & 0.080 & 0.110 & 0.109 & 0.913 & 0.934
\\
$\beta_{2}$ & -0.6 & -0.692 & 0.083 & -0.690 & 0.083 & 0.105 & 0.103 & 0.942 & 0.952
\\
$\beta_{3}$ & -0.6 & -0.696 & 0.083 & -0.692 & 0.083 & 0.107 & 0.107 & 0.933 & 0.925
\\
$\beta_{4}$ & -0.5 & -0.712 & 0.184 & -0.718 & 0.186 & 0.166 & 0.164 & 0.712 & 0.717
\\
$\beta_{5}$ & -0.5 & -0.572 & 0.165 & -0.576 & 0.166 & 0.181 & 0.179 & 0.952 & 0.943
\\
$\gamma_{0}$ & 0.4 & 0.339 & 0.058 & 0.338 & 0.058 & 0.060 & 0.060 & 0.827 & 0.811
\\
$\gamma_{1}$ & 0.3 & 0.306 & 0.020 & 0.307 & 0.021 & 0.020 & 0.020 & 0.913 & 0.906
\\
$\gamma_{2}$ & 0.3 & 0.307 & 0.020 & 0.307 & 0.020 & 0.020 & 0.021 & 0.904 & 0.906
\\
$\gamma_{3}$ & 0.3 & 0.309 & 0.020 & 0.310 & 0.021 & 0.020 & 0.020 & 0.900 & 0.901
\\
$\gamma_{4}$ & -0.3 & -0.315 & 0.035 & -0.317 & 0.034 & 0.036 & 0.036 & 0.933 & 0.925
\\
$\gamma_{5}$ & 0.6 & 0.625 & 0.036 & 0.625 & 0.035 & 0.045 & 0.045 & 0.942 & 0.962  \\
\bottomrule
\label{tab:large sample estimation(40 percent)}
\end{tabular}
\end{center}
\end{table} 

\begin{table}[htbp] 
\caption{Average number of basis functions and iterations for a $40\%$ zero-inflation proportion and a moderate spatial correlation ($c = 0.3$) at $n = 3000$ across 200 replicates. Standard deviations are provided in parentheses. ``AIC'' represents the number of basis functions selected by Akaike's information criterion, while ``Fix'' indicates a fixed number of basis functions.
}
\begin{center}
\begin{tabular}{cccc} \toprule
& $\hat{K}_{1}$ & $\hat{K}_{2}$ & Number of iterations
\\
\cmidrule(r){2-3} \cmidrule(r){4-4}
\multirow{2}*{AIC} & 23.00  & 22.00  & 6.18 \\
			~      & (1.10) & (0.82) & (0.89) \\
\multirow{2}*{Fix} & 30.00  & 30.00  & 6.08 \\
			     ~ & (0.00) & (0.00) & (0.63)
\\
\bottomrule
\end{tabular}
\end{center}
\label{tab:number of basis functions and iteration for large sample(40 percent)}
\end{table} 

\section{Application to Taiwan's daily rainfall data}\label{sec6:application}
 In this section, we apply the flexible spatial zero-inflated count model to analyze Taiwan's daily rainfall data from 2016, obtained from the Taiwan Climate Change Projection Information and Adaptation Knowledge Platform (TCCIP). The objective is to investigate the relationship between extreme rainfall events and potential contributing factors. The dataset includes rainfall observations from 1,269 distinct grid points across Taiwan. To identify extreme rainfall events, we adopt the rainfall magnitude standards provided by the Central Weather Bureau (CWB) of Taiwan, defining extreme events as those where the daily cumulative rainfall surpasses 350 millimeters.

\begin{figure}[htbp]\centering
 \begin{center}
  \includegraphics[scale=0.5]{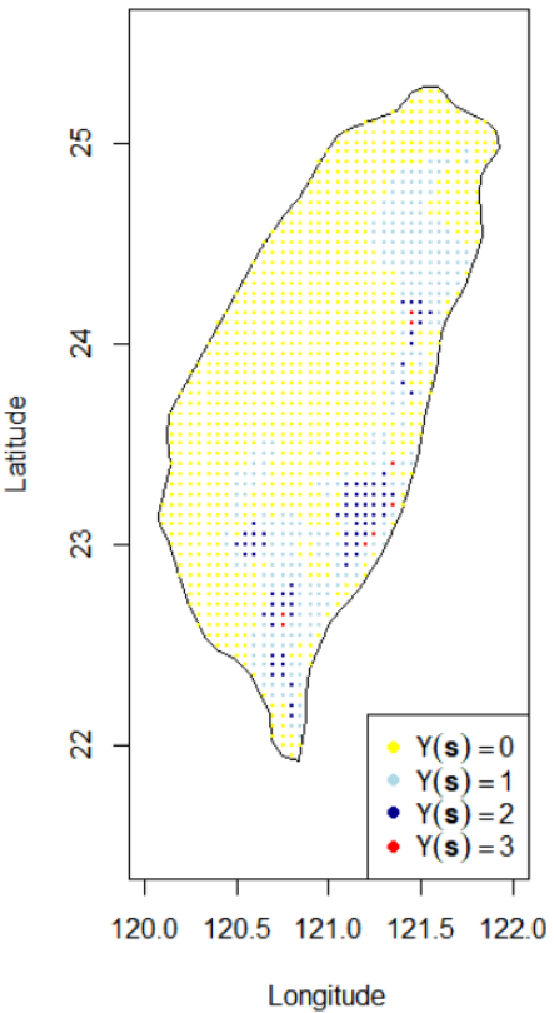}
   \caption{The spatial distribution of the number of extreme rainfall occurrences across Taiwan in 2016.}
  \label{fig:pattern of extreme rainfall across taiwan}
  \end{center}
 \end{figure}

 In this dataset, the response variable $Y(\bm{s}_{i})$ represents the number of extreme rainfall events recorded at each location $\bm{s}_{i}$, where $i = 1, 2, \ldots, 1269$. To explore the factors affecting the occurrence probability $\phi(\bm{s}_{i})$ and prevalence intensity $\lambda(\bm{s}_{i})$ of extreme rainfall events, we adopt the following five meteorological and geographical variables:
 \begin{itemize}
  \item Daily maximum temperature (MT)
  \item Daily average temperature (AT)
  \item Relative humidity (Hum)
  \item Atmospheric pressure (Pre)
  \item Altitude (Alt)
 \end{itemize}
 Due to the differences in measurement units, all variables are standardized prior to analysis. The zero-inflation proportion of the response variables in this dataset is approximately $72\%$. Figure \ref{fig:pattern of extreme rainfall across taiwan} illustrates the spatial distribution of $Y(\bm{s}_i)$ across Taiwan for $i = 1, \dots, 1269$ in 2016. We then model $\phi(\bm{s}_{i})$ and $\lambda(\bm{s}_{i})$ using the equations specified in \eqref{phi} and \eqref{lambda}, where $\bm{v}(\bm{s}_{i})$ and $\bm{u}(\bm{s}_{i})$ represent all variables \{MT, AT, Hum, Pre, Alt\}, and $\bm{\beta}$ and $\bm{\gamma}$ are the corresponding vectors of regression coefficients associated with these variables.

  In this model, the regression coefficients are estimated using the proposed iterative estimation algorithm outlined in Section \ref{sec4.3:algorithm}, resulting in the estimated occurrence probabilities, $\hat{\phi}(\bm{s})$, and the estimated prevalence intensity, $\hat{\lambda}(\bm{s})$, as follows:
   \begin{align}\label{estimated.phi}
    \hat{\phi}(\bm{s})=\displaystyle\frac{\exp\left(-19.13+40.4{\rm{MT}}(\bm{s})-83.57{\rm{AT}}(\bm{s})-3.99{\rm{Hum}}(\bm{s})-0.04{\rm{Pre}}(\bm{s})-56.23{\rm{Alt}}(\bm{s})\right)}
    {1+\exp\left(-19.13+40.4{\rm{MT}}(\bm{s})-83.57{\rm{AT}}(\bm{s})-3.99{\rm{Hum}}(\bm{s})-0.04{\rm{Pre}}(\bm{s})-56.23{\rm{Alt}}(\bm{s})\right)}
   \end{align} 
   and
   \begin{align}\label{estimated.lambda}
    \hat{\lambda}(\bm{s})=\exp\left(-1.63+0.61{\rm{MT}}(\bm{s})-0.2{\rm{AT}}(\bm{s})-0.01{\rm{Hum}}(\bm{s})+0.12{\rm{Pre}}(\bm{s})+0.75{\rm{Alt}}(\bm{s})\right).
   \end{align} 

\begin{figure}\centering
 \begin{center}
  \includegraphics[scale=0.5]{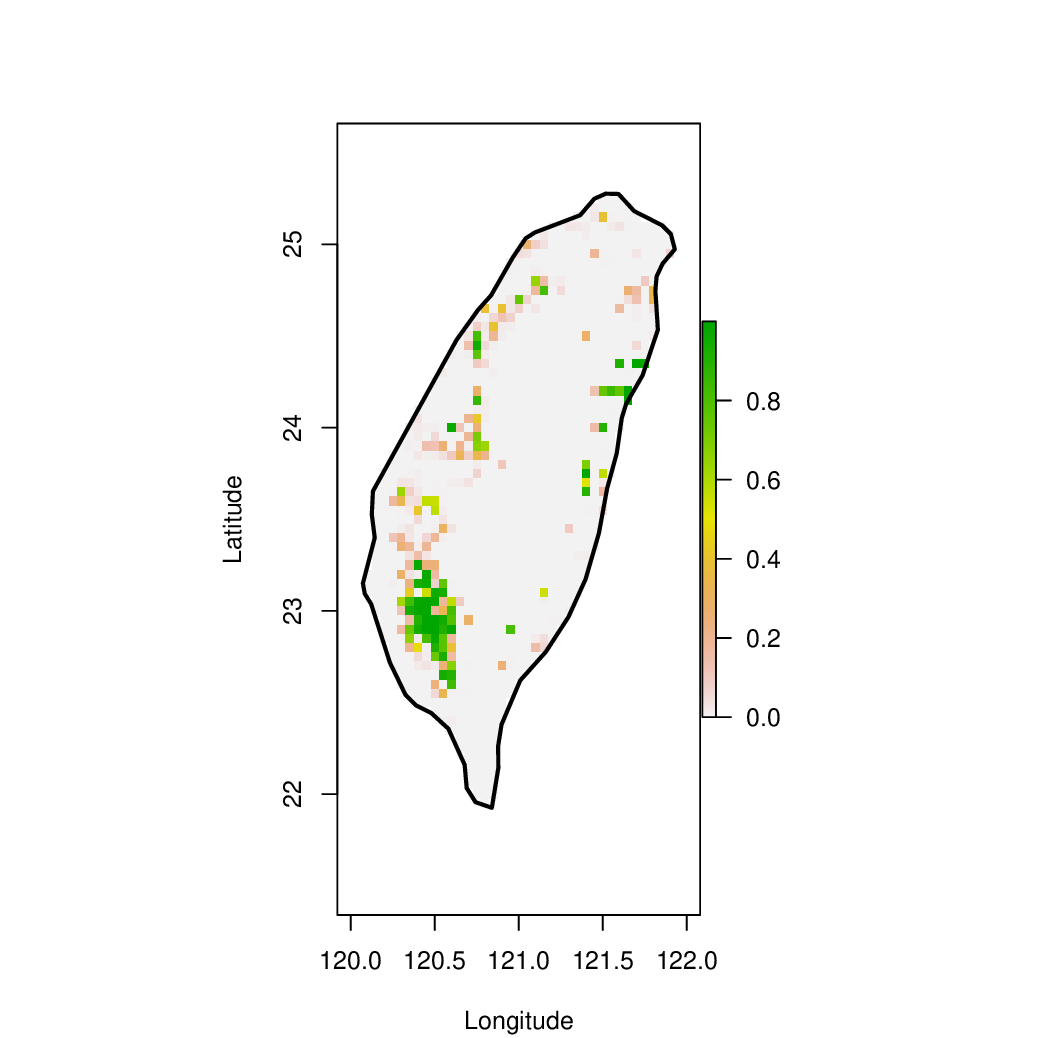}
  \hspace{-1.5cm}
  \includegraphics[scale=0.5]{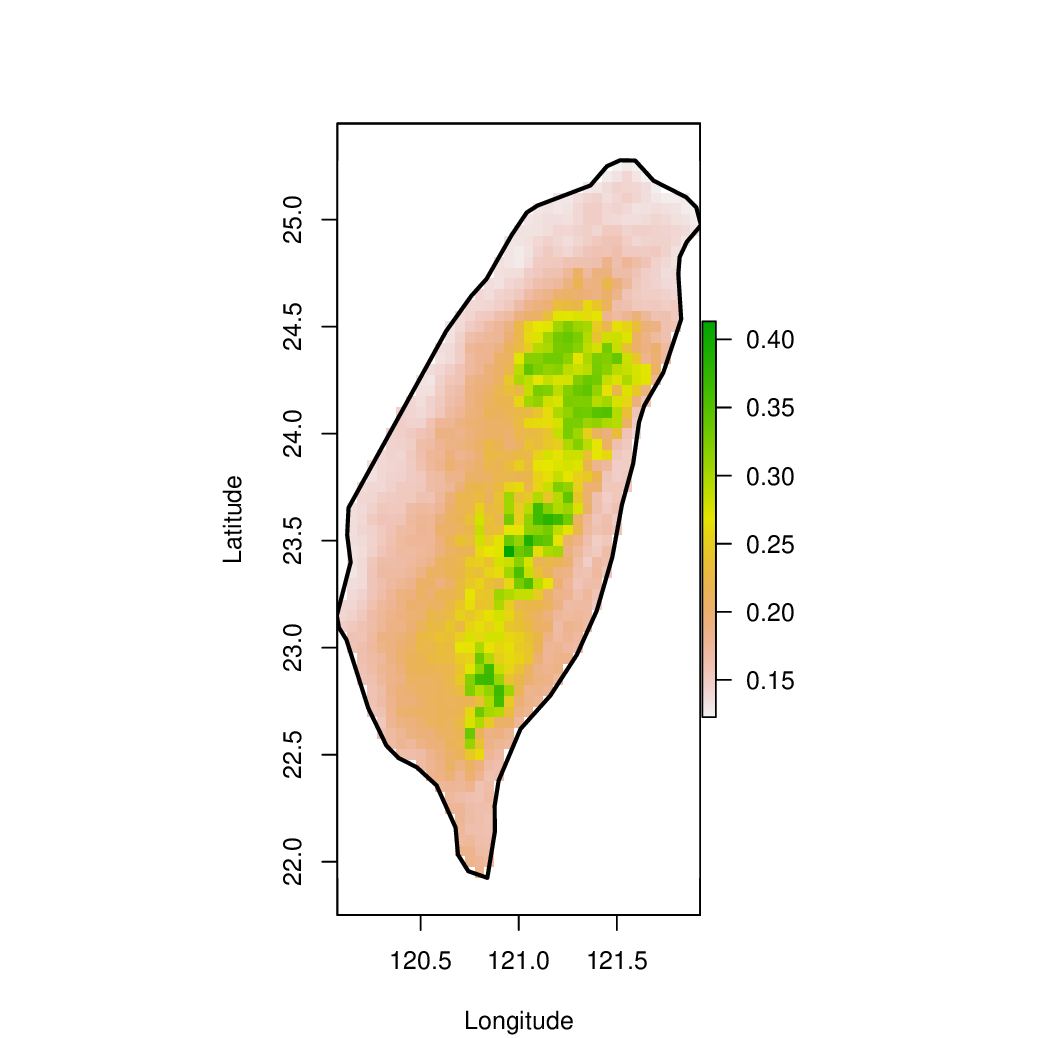}
   \caption{Estimated surfaces of $\phi(\bm{s})$ (Left) and $\lambda(\bm{s})$ (Right) across Taiwan in 2016.}
  \label{fig: phi.lambda}
  \end{center}
 \end{figure}

 According to (\ref{estimated.phi}) and (\ref{estimated.lambda}), Figure \ref{fig: phi.lambda} illustrates the spatial distribution patterns of the occurrence probability $\phi(\bm{s})$ and the prevalence intensity $\lambda(\bm{s})$ for extreme rainfall events across Taiwan. As shown on the left side of Figure \ref{fig: phi.lambda}, the estimated occurrence probabilities $\hat{\phi}(\bm{s})$ are notably higher across much of western Taiwan, indicating a lower propensity for extreme rainfall events in these regions. In contrast, the estimated occurrence probabilities $\hat{\phi}(\bm{s})$ are significantly lower along the Central Mountain Range of Taiwan, suggesting a higher risk of extreme rainfall in these areas. This spatial distribution pattern aligns with meteorological observations, highlighting the influence of orographic effects and prevailing weather patterns in increasing the likelihood of intense precipitation in these regions.

 On the right side of Figure \ref{fig: phi.lambda}, the prevalence intensity $\lambda(\bm{s})$ is higher in high-altitude regions, which is consistent with the established understanding that orographic lifting plays a significant role in enhancing rainfall intensity in mountainous areas. The congruence between the modeled spatial trends and observed meteorological conditions underscores the robustness and applicability of the proposed flexible spatial zero-inflated model for analyzing extreme rainfall events. This model offers valuable insights for understanding the spatial dynamics of extreme precipitation, which is critical for effective climate risk management and adaptation strategies in Taiwan.

 In this analysis, the numbers of basis functions selected were $\hat{K}_1=343$ and $\hat{K}_2=329$, which are significantly smaller than $n=1269$. This reduction in dimensionality enhances computational efficiency while maintaining the ability to accurately model spatial patterns. While $n=1269$ may not yet pose significant challenges for existing spatial models, it is important to note that computing and storing an $n^2$-dimensional covariance matrix can become increasingly burdensome as $n$ grows larger. Our method demonstrates how basis function expansions can effectively address these computational demands, offering a scalable solution for analyzing larger datasets without sacrificing precision.

 In the dataset used for this analysis, the observed response values are limited to $Y(\bm{s}) \in\{0, 1, 2, 3\}$, reflecting relatively low counts. Despite this limitation, the proposed GEE-based method remains robust and effective due to its reliance on moment-based inference rather than explicit distributional assumptions. This analysis further underscores the advantages of the GEE-based approach in effectively handling low-count data scenarios. By not imposing rigid distributional assumptions, the proposed method demonstrates greater flexibility and robustness, making it a valuable tool for analyzing spatial zero-inflated count data with limited count ranges.
\section{Conclusion and discussion}\label{sec7:conclusion and discussion}
 In this study, we introduced a novel approach for constructing spatial structures within the spatial zero-inflated count model, utilizing basis functions derived from thin-plate splines. This method enables a flexible and computationally efficient representation of the spatial covariance matrix, effectively capturing spatial correlations in the data. Our projection-based approach addresses the challenges associated with high-dimensional matrices and enhances the efficiency of parameter estimation within the generalized estimating equations (GEE) framework. By iteratively approximating the true spatial covariance structure during model estimation, our method significantly accelerates convergence to the true regression parameters and overcomes the computational difficulties linked to high-dimensional matrix inversion. The robustness and practicality of our approach were demonstrated through extensive numerical simulations and an application to Taiwan's daily rainfall data for 2016.

 Recently, the method proposed by Lee and Haran (2024) employs a Bayesian framework with PICAR basis expansion to estimate parameters in large spatial zero-inflated count data models. While Bayesian-based methods are effective in many cases, they rely heavily on strong distributional assumptions about the observations. However, in practical settings, the true distribution of the observations is often unknown, especially when spatial correlation is present. This reliance on assumed distributions can make Bayesian-based methods sensitive to deviations from these assumptions, potentially reducing their robustness in practical applications. In contrast, our proposed GEE-based model only requires assumptions about the means of the responses (e.g., Equations \eqref{zero mean} and \eqref{nonzero mean}) without relying on full distributional assumptions. This fundamental distinction makes the GEE-based approach more flexible and better suited for applications where the true data-generating process is uncertain or when robust inference is critical. Furthermore, Shen and Chen (2024) demonstrated through simulation studies that likelihood-based methods are sensitive to the underlying data structure. By comparison, the GEE-based method, which relies on moment-based inference instead of complete distributional specifications, exhibits greater robustness across various data scenarios, offering a distinct advantage over Bayesian-based methods such as PICAR (Lee and Haran 2024)
 
 This study extends the GEE framework of Shen and Chen (2024) by incorporating a more comprehensive covariance structure, enhancing its scalability and applicability to large-scale spatial zero-inflated count data. This advancement effectively addresses the computational challenges associated with large datasets while maintaining the flexibility and robustness of the GEE approach. A significant improvement in computational efficiency is achieved through our parameter estimation technique, which leverages the GEE framework with an estimated covariance matrix. Results from various simulation scenarios confirm the effectiveness and reliability of our method. Another key advantage of our approach lies in its flexibility, as it does not require specifying a parametric model to construct spatial structures, making it highly applicable to practical scenarios where the true data-generating process may be unknown or complex. While the Newton-Raphson method provides a robust framework for parameter estimation, careful selection of the convergence criteria and the maximum number of iterations is essential to ensure reliable and efficient results.
 
 Moreover, spatial confounding, which occurs when covariates and spatial random effects are collinear, is a well-documented challenge in spatial modeling (Hodges and Reich 2010; Paciorek 2010). This phenomenon can bias parameter estimates and complicate the interpretation of covariate effects in frameworks that combine covariates with smooth spatial errors. In our study, we mitigate the potential effects of spatial confounding by adopting the GEE approach, which focuses on the marginal mean and covariance structure of the data rather than fully specified random effect models. This marginal approach reduces the direct interaction between covariates and spatial random effects, minimizing the risk of confounding. Additionally, our model employs a flexible TPS-based covariance structure to effectively capture spatial dependence without imposing overly rigid assumptions that might exacerbate confounding. However, we acknowledge that spatial confounding is not explicitly addressed within our framework. While the GEE method’s moment-based inference offers robustness against certain modeling assumptions, it does not fully eliminate confounding. Further research could explore incorporating confounding-robust basis function selection or spatial adjustment techniques, such as restricted spatial regression (Reich et al. 2006; Chiou et al. 2019; Khan and Calder 2020) and related methods (Hughes and Haran 2013; Dupont et al. 2022; Chiou and Chen 2025) into the TPS framework. By addressing this limitation, we aim to inspire future advancements in refining the proposed model for broader applications.

 Despite the advancements presented in this paper, a notable gap remains in the field: there is currently no established method for selecting significant covariates for the occurrence probability and prevalence intensity in spatial zero-inflated count models, particularly when the data's underlying distribution is unknown. Future research should focus on developing a model selection criterion specifically tailored to the flexible spatial zero-inflated count model. This would involve creating a distribution-free variable selection criterion and devising an efficient algorithm to identify significant covariates. Addressing these challenges will further enhance the utility of these models in a wide range of statistical analyses, contributing to more robust and reliable inferences in the study of spatially correlated count data. We leave these issues for future researches.


\section*{Declarations}
 No potential conflict of interest was reported by the authors.

\section*{\textsf{References}}
\begin{description}
 \item Adegboye OA, Leung DH, Wang YG~(2018) Analysis of spatial data with a nested correlation structure. J R Stat Soc Ser C-Appl Stat 67:329-354.

 \item Agarwal DK, Gelfand AE, Citron-Pousty S~(2002) Zero-inflated models with application to spatial count data. Environ Ecol Stat 9:341-355.

 \item Banerjee S, Gelfand AE, Finley AO, Sang H~(2008) Gaussian predictive process models for large spatial data sets. J R Stat Soc Ser B-Stat Methodol 70:825-848.

 \item B{\"o}hning D, Dietz E, Schlattmann P, Mendonga L, Kirchner U~(1999) The zero-inflated Poisson model and the decayed, missing and filled teeth index in dental epidemiology. J R Stat Soc Ser A-Stat Soc 162:195-209.
 
 \item Chen CWS, Chen CS~(2024) Spatial-temporal hurdle model vs. spatial zero-inflated GARCH model: analysis of weekly dengue fever cases. Stoch Environ Res Risk Assess 38:2119-2134.
 
 \item Chiou YH, Chen CS~(2025) A frequentist approach on fixed effects estimation for spatially confounded regression models. Environ Ecol Stat 32:523-555.
     
 \item Chiou YH, Yang HD, Chen CS~(2019) An adjusted parameter estimation for spatial regression with spatial confounding. Stoch Environ Res Risk Assess 33:1535-1551

 \item Cressie N~(1993) \textit{Statistics for Spatial Data} (revised edition). New York: Wiley.

 \item Cressie N, Johannesson G~(2008) Fixed rank kriging for very large spatial data sets. J R Stat Soc Ser B-Stat Methodol 70:209-226.

 \item Duchon J~(1977) Splines minimizing rotation-invariant semi-norms in Sobolev spaces. In: Schempp, W., Zeller, K. (eds) Constructive Theory of Functions of Several Variables. Lecture Notes in Mathematics, vol 571. Springer, Berlin, Heidelberg.
     
 \item Dupont E, Wood SN, Augustin NH~(2022) Spatial+: a novel approach to spatial confounding.
        Biometrics 78:1279-1290.
     
 \item Farewell V T, Sprott DA~(1988) The use of a mixture model in the analysis of count data. Biometrics 44:1191-1194.

 \item Green PJ, Silverman BW~(1993) \textit{Nonparametric Regression and Generalized Linear Models: A Roughness Penalty Approach}. Boca Raton: CRC Press.

 \item Guan Y, Haran M~(2018)  A computationally efficient projection-based approach for spatial generalized linear mixed models. J Comput Graph Stat 27:701-714.

 \item Hall DB, Zhang Z~(2004) Marginal models for zero inflated clustered data. Stat Model 4:161-180.

 \item Harville DA~(1997) \textit{Matrix Algebra From a Statistician's Perspective}. New York: Springer.

 \item He H, Wang WJ, Hu J, Gallop R, Crits-Christoph P, Xia YL~(2015) Distribution-free inference of zero-inflated binomial data for longitudinal studies. J Appl Stat 42:2203-2219.
     
 \item Hodges JS, Reich BJ~(2010) Adding spatially-correlated errors can mess up the fixed effect you love.
       Am Stat 64:325-334.
       
 \item Hughes J, Haran M~(2013) Dimension reduction and alleviation of confounding for spatial generalized linear mixed models. J R Stat Soc Ser B-Stat Methodol 75:139-159.

 \item Katzfuss M~(2017) A multi-resolution approximation for massive spatial datasets. J Am Stat Assoc 112:201-214.
     
 \item Khan K, Calder CA~(2020) Restricted spatial regression methods: implications for inference. J Am Stat Assoc 117:482–494.

 \item Lambert D~(1992) Zero-inflated Poisson regression, with an application to defects in manufacturing. Technometrics 34:1-14.

 \item Lee BS, Haran M~(2024) A class of models for large zero-inflated spatial data. J Agric Biol Environ Stat https://doi.org/10.1007/s13253-024-00619-9
 \item Lee CE, Kim S~(2017) Applicability of zero-inflated models to fit the torrential rainfall count data with extra zeros in South Korea. Water 9:123.

 \item Liang KY, Zeger SL~(1986) Longitudinal data analysis using generalized linear models. Biometrika 73:13-22.

 \item McCullagh P, Nelder JA~(1989) \textit{Generalized Linear Models}. Boca Raton: Chapman and Hall/CRC.
 
 \item Neelon B, Ghosh P, Loebs PF~(2012) A spatial Poisson hurdle model for exploring geographic variation in emergency department visits. J R Stat Soc Ser A-Stat Soc 176:389-413.

 \item Neelon B, Zhu L, Neelon SEB~(2015) Bayesian two-part spatial models for semicontinuous data with application to emergency department expenditures. J R Stat Soc Ser A-Stat Soc 16:465-479.

 \item Nychka D, Bandyopadhyay S, Hammerling D, Lindgren F, Sain S~(2015) A multiresolution Gaussian process model for the analysis of large spatial datasets. J Comput Graph Stat 24:579-599.

 \item Paciorek CJ~(2010) The importance of scale for spatial-confounding bias and precision of spatial
       regression estimators. Stat Sci 25:107-125.
       
 \item Park J, Haran M~(2020) Reduced-dimensional Monte Carlo maximum likelihood for latent Gaussian random field models. J Comput Graph Stat 30:69-283.

 \item Rathbun SL, Fei S~(2006)  A spatial zero-inflated poisson regression model for oak regeneration. Environ Ecol Stat 13:409-426.
     
 \item Reich BJ, Hodges JS, Zadnik V~(2006) Effects of residual smoothing on the posterior of the fixed effects in disease-mapping models. Biometrics 62:1197-1206.

 \item Shen CW, Chen CS~(2024) Estimation and selection for spatial zero-inflated count models. Environmetrics 35:e2847.
     
 \item Sherman M~(2011) \textit{Spatial Statistics and Spatio-temporal Data}. Chichester: Wiley.
 
 \item Silva AT, Portela MM, Naghettini M~(2014) On peaks-over-threshold modeling of floods with zero-inflated Poisson arrivals under stationarity and nonstationarity. Stoch Environ Res Risk Assess 28:1587–1599.
     
 \item Taiwan Climate Change Projection and Information Platform (TCCIP)~(2023). Gridded observation data. https://tccip.ncdr.nat.gov.tw/ds\_03\_eng.aspx.

 \item Tzeng SL, Huang HC~(2018) Resolution adaptive fixed rank kriging. Technometrics 60:198-208.

 \item Wahba G~(1985) A comparison of GCV and GML for choosing the smoothing parameter in the generalized spline smoothing problem. Ann Stat 13:1378-1402.

 \item Wahba G~(1990) {\it Spline Models for Observational Data}. Philadelphia: Society for Industrial and Applied Mathematics.

 \item Wikle CK~(2010) Low-Rank Representations for Spatial Processes. In \textit{Handbook of Spatial Statistics}, eds. A. E. Gelfand, P. J. Diggle, M. Fuentes, and P. Guttorp, Boca Raton, FL: CRC Press, 107-118.
     
 \item Wood SN~(2017) \textit{Generalized Additive Models: An Introduction with R. 2nd ed.}. New York: Chapman and Hall/CRC.

\end{description}
\end{document}